# Do cooperative cycles of hydrogen bonding exist in proteins?

John N. Sharley, University of Adelaide. arXiv:1601.01792

john.sharley@pobox.com

## Table of Contents

## 1   Abstract

The closure of cooperative chains of Hydrogen Bonding, HB, to form cycles can enhance cooperativity [1]. Cycles of charge transfer can balance charge into and out of every site, eliminating the charge build-up that limits the cooperativity of open unidirectional cooperative chains. If cycles of cooperative HB exist in proteins, these could be expected to be significant in protein structure and function in ways described below. We find no mention of an example of this kind of cycle in the literature. We investigate whether cooperative HB cycles not traversing solvent, ligand or modified residues occur in proteins by means including search of Nuclear Magnetic Resonance [2], NMR, spectroscopy entries of the Protein Data Bank [3], PDB.

For the direct interactions of inter-amide HB, when the energy associated with Natural Bond Orbital [1], NBO, steric exchange is deducted from that of NBO donor-acceptor interactions, the result is close to zero, so that HB is not primarily due to the sum of direct inter-amide NBO interactions. The NBO binding energy is primarily associated with the increase in resonance of the amides, a consequence of which is that the majority of the NBO binding energy is susceptible to variation by electrostatic field with component parallel or antiparallel to an amide C-N bond [4].

The question of what geometry most favours HB in amides is revisited with emphasis on the inequivalence of amide/carbonyl oxygen lone pairs.

A possible avenue for the design of HB-chaining polymers with improved stability is discussed.

## 2 Introduction

### 2.1 Review of Resonance-Assisted HB

The hydrogen atom is unique in that it has no core electronic shell. A donor orbital can overlap most of the H of an H-X antibonding orbital rather than being limited to the region outside of a nodal boundary [1]. A lone pair of electrons is particularly suitable as donor in this donor-acceptor interaction. An unusually strong donor-acceptor interaction, the HB, arises in this manner. That the HB is primarily resonance-type covalency or charge transfer rather than electrostatic in nature is most recently supported by evidence of anti-electrostatic HB [5].

An HB, nominated HB1, induces repolarization of the H-X acceptor antibonding orbital and its corresponding bonding orbital, resulting in elevated partial negative charge on atom X. Lone pair orbitals on the X atom become of higher energy and more diffuse. If one of the X atom lone pair orbitals donates charge for another HB, HB2, then HB2 will be of greater binding energy than in the absence of the repolarization of H-X induced by HB1. HB1 is also of greater binding energy than in the absence of HB2, since HB2 transfers charge from X allowing further repolarization of H-X and better overlap with the HB1 donor. The charge transfer, CT, of both of HB1 and HB2 is greater when both exist, that is, they are cooperative.

Resonance-Assisted HB [6, 7], RAHB, occurs when resonance gives the lone pair donor anionic character or the acceptor orbital cationic character. RAHB differs from Charge Assisted HB [1], CAHB, in that the assisting charge of CAHB is not varied by resonance. The resonance of RAHB is greater again where the resonant group can both accept and donate hydrogen bonds and both hydrogen bonds are present. In the case of the amide group, highly cooperative HB results. Amide resonance features in RAHB protein secondary structures such as alpha helices [8] and beta sheets [9].

A donor-acceptor interaction is a partial CT from a donor orbital to an acceptor orbital. The donor assumes a more cationic character and the acceptor a more anionic character as a result of the CT, which tends to oppose further CT. In an arrangement of donor-acceptor interactions such that the CT from a site is balanced by other CT to the site, the magnitude of the CT is not limited by charge imbalance [10]. In an open chain of CT between similar units at similar successive orientation, the charge imbalance is lower closer the middle of the chain for it is there that CT more closely balances charge at each site, and so CT peaks there. When the chain of CT is closed, the charge is balanced at all sites and the magnitude of each CT is the same. Unless varied by the necessary geometry change between open and closed conformations of chain, each CT in the closed conformation is at least that of the peak of the open conformation. Unless the number of units in a chain is such that an asymptotic limit of cooperativity has been closely approached, each CT in the closed conformation will exceed rather than equal that of the peak of the open conformation. With reference to hydrogen fluoride clusters, Weinhold and Landis [1] remark that "the strong preference for cyclic clusters is quite perplexing from a classical dipole-dipole viewpoint". In investigating cyclic cooperativity in proteins, we take the NBO view that HB has partially covalent nature. This nature is most pronounced when charge transfer is largest, which tends to be when the HB is short.

## 2.2   Phenomena anticipated if cyclic cooperative HB exists in proteins

If cyclic cooperative HBs were to exist in proteins, the following phenomena might be anticipated, further motivating study of the possibility that such cycles exist.

If backbone amide HB chains in secondary structure were connected by additional cooperative units such that cooperative cyclic HB existed, additional stabilization of the secondary structures involved would result, since the sum of free energy of HB in the secondary structure part of the cycle would be higher. Various means of cooperatively cyclizing secondary structure HB chains are conceivable. For example, one spine from each of multiple alpha helices might be connected into a cycle by sidechain amides. An alpha helix might participate in three such arrangements, one for each of its spines. A beta sheet might have pairs of HB chains adjacent in the sheet connected at each end to form cycles, one cycle per pair of HB chains. A beta sheet might have its HB chains connected to form a longer chain, with a spine of an alpha helix diagonally across the sheet to form a cycle.

Binding specificity might be increased by a cooperative cycle which passes through both binding partners. This difference in binding energy between the complete cycle and incomplete cycle would give increased binding specificity. If the cycle traversed the binding interface more than twice, further specificity would result.

A cooperative cycle might be viewed as a spatially distributed store of binding energy. If energy is supplied to break the cycle at one point, the binding energy of all HBs in the cycle is decreased. Closing the cycle increases the binding energy of all HBs in the cycle. This is not dependent on the number of units in the cycle provided that number is under a unit limit of cooperativity. We showed that estimations of this asymptotic limit in beta sheet using established DFT methods and a range of basis sets must be set aside [11]. The making and breaking of cycles may result in allostery [12]. Two cooperative cycles might be mutually exclusive, with each associated with a conformation.

Some molecular chaperones [13] or other binding partners might form cooperative cycles which include cooperative units in their client, changing conformation in the client, and upon input of energy such as from hydrolyzing ATP [14], break this cooperative cycle in which they are involved, allowing formation of a cooperative cycle internal to the client.

Cooperative cycles might enhance the stability of amyloid fibrils [15] which have long cross-strand chains of inter-peptide HBs. The stacking of beta sheets might give more opportunities for closure of cycles, with sidechains completing cycles with backbone amide chains in the adjacent sheets.

## 3  Notation

"->" denotes NBO resonance-type charge transfer and "|" denotes NBO steric exchange repulsion. "(" and ")" enclose specification of an orbital type and follow an atom name for single-center NBOs and a pair of atom names separated by "-" for two-center NBOs.

Examples: N(lp) for the amide nitrogen lone pair NBO, O(lp-p) for the oxygen p-type lone pair NBO, O(lp-s) for the s-rich lone pair NBO, C-O(p)* for the pi carbonyl antibonding orbital NBO and N(lp)->C-O(p)* for the primary amide resonance type charge transfer.

## 4  Methods

### 4.1  Counterpoise correction

Mentel and Baerends [16] found that the use of the Counterpoise Correction [17] for Basis Set Superposition Error [18, 19], BSSE, was not justifiable. In accordance with this finding, we do not use this correction in these experiments.

### 4.2  Dispersion correction

We found that the D3 correction [20, 21] decreased amide carbonyl sigma/pi separation with the three method tested. Since we are primarily concerned with resonance, we broadly avoid this correction, using it only for comparison in one experiment.

### 4.3 Software packages

Methods used in experiments are as implemented by Gaussian 09 D.01 [22], Orca 3.0.3 [23-25] and TeraChem 1.5K [20, 21, 26, 27]. Unless otherwise stated, default grids and optimization and SCF convergence limits were used, except that the Orca option VeryTightSCF was used throughout as were cartesian coordinates for geometry optimization with TeraChem.

A pre-release version of NBO [28] was used for its XML [29] output option. The XML was queried with XQuery 3.0 [30] or XSLT 3.0 [31] as implemented by Saxon-PE 9.6.0.4 [32], and the results imported into Excel 2013 [33].

Jmol 14.2.2_2014.06.29 [34] was used for visualization of orbitals.

Molecular coordinates are depicted by UCSF Chimera 1.10.2 [35].

### 4.4 Haskell

As detection of cooperative hydrogen bonding cycles in the PDB is dependent here on program correctness without other confirmation except when a potential example is flagged for investigation by quantum chemical means, some emphasis was placed on high probability of this correctness. The programming language Haskell [36] was used in an attempt to address the problem of program errors remaining undiscovered in all scientific codes. It is common that errors in scientific codes are discovered long after calculations have been performed by such codes. This problem increases with the size of the code base, and scientific programs tend to become large. This problem will not be fixed until scientific codes are formally proven [37] to solve highest-level equations, but until then pure functional languages [38] such as Haskell represent progress which is practicable. In pure functional languages, there are no variables, merely labels immutably bound to the results of function evaluation. By default, the programmer does not control the flow of program execution, and execution follows the necessary data dependencies. Haskell users often remark that if program code passes the compiler checks, it is likely right first time [39]. In summary, Haskell was used in an endeavour to improve reliability of results beyond that likely to be achievable with the most diligent use of imperative languages [40].

Detection of cycles was first implemented here in Haskell without a list comprehension [41], but the results were sufficiently surprising that a simpler list comprehension implementation was written. The different implementations returned the same results.

## 5 Results and Discussion

### 5.1 HB angle

#### 5.1.1 Experiments involving hydrogen bonding between an amide group and hydrogen fluoride

It has long been appreciated that C-O..H-N linearity is not optimal for amide-amide hydrogen bonding, though this is usually ascribed to carbonyl lone pairs being equivalent sp2 hybrids having trigonal planar geometry [42]. However, these lone pairs are far from equivalent. As for water oxygen lone pairs when not engaged in intermolecular bonding [43, 44], carbonyl oxygen lone pairs are distinctly inequivalent, but unlike water, are exceedingly reluctant to become more equivalent when engaged in bonding. The morphology of the amide oxygen lone pair NBOs is shown in Figure 1 and Figure 2. This electron density is not equivalent to two similar hybrids, and NBOs are not unitarily equivalent to canonical molecular orbitals [43, 44]. Substantial maintenance of inequivalence for the case of Hydrogen Fluoride HB with N-methylformamide oxygen at given C-O-F angles in the amide plane with F distal to N as in Figure 3 is shown in Ap1:Figure 26 and Ap1:Figure 28.

With hydrogen fluoride as a probe of HB with N-methylformamide oxygen, and with the C-O-F angle constrained to given angles in the amide plane distal to N, a range of observations are made. Angles are given as deviation from linear, so that collinear C-O-F is given as 0 degrees rather than 180 degrees. This angle corresponds to the angle between the vectors C-O and O-F. Figure 4, Figure 5 and Ap1:Figure 23 contrast observations of H-F* NBO occupancy, HB length and C-O(p)* NBO occupancy with and without the extra constraint that the C-O-H (H of HF) angle is the same as the C-O-F angle. To the degree precision shown, the maximum H-F* occupancy occurs at 75 degrees when the C-O-H constraint is used and 80 degrees when it is not. HB length minimum in the 0 to 90 degree range considered occurs at 70 degrees with the C-O-H constraint and 75 degrees without. The maximum HB length is seen at 0 degrees i.e. with collinear C-O..H-F. The maximum amide resonance as given by C-O(p)* NBO occupancy occurs at 85 degrees with the C-O-H constraint and 75 degrees without. The minimum amide resonance occurs at 5 degrees with the C-O-H constraint and 0 degrees without.

Further figures refer to the case without C-O-H constraint. Ap1:Figure 24 shows that the H-F* NBO occupancy is similarly calculated by SCS-MP2/aug-cc-pVTZ and DLPNO-CCSD(T)/aug-cc-pVTZ with coulomb and correlation auxiliary basis sets at SCS-MP2/aug-cc-pVTZ optimized geometry, indicating these occupancy figures do not arise of a unique property of SCS-MP2.

Ap1:Figure 25 shows the F-H-O angle at given C-O-F angle with 4 wavefunction methods each with 2 correlation consistent basis sets. There is consensus at C-O-F angle of 55 degrees that the F-H-O angle is zero i.e. F, H and O collinear, and less consensus that at C-O-F angle of 0 degrees that the F-H-O angle is again zero.

Ap1:Figure 26, Ap1:Figure 27 and Ap1:Figure 28 show the fraction of p character of O(lp-p), O(lp-s), C-O(p) and C-O(s) at varying C-O-F angles at both SCS-MP2 and DLPNO-CCSD(T). The p character of C-O(p) does not vary, C-O(s) varies by ~0.5 percent, O(lp-p) and O(lp-s) vary by ~1.6 percent, so that O(lp-p) and O(lp-s) remain largely inequivalent.

Ap1:Figure 29 shows the variation of NBO donor-acceptor Second-Order Perturbation Theory [1], SOPT, and NBO Steric Exchange Energy [1] between the amide oxygen lone pair NBOs and the H-F and H-F* NBOs at C-O-F angle. For a balanced view of energetics of an interaction, donor-acceptor interactions must be considered against steric interaction of the donor and the acceptor's associated bonding orbital e.g. O(lp-p)->H-F* and O(lp-p)|H-F must be considered together. Notable features are that O(lp-p)->H-F* minus O(lp-p)|H-F, referred to here as p delta, exceeds O(lp-s)->H-F* minus O(lp-s)|H-F, referred to here as s delta, meaning the energy gradient is determined by the p-type lone pair. The s delta declines only slightly with C-O-F angle. At 75 degrees the p delta is 13.08 kcal/mol and the s delta is 2.98 kcal/mol.

Figure 6 shows the p delta plus s delta for a range of DFT methods at SCS-MP2/aug-cc-pVTZ optimized geometry. Ap1:Figure 30 shows this value when the geometry optimization is also done at each method, though due to Gaussian's aversion to near-linear ModRedundant constraints, the range of angles starts at 15 degrees. Ap1:Figure 31 shows this value for LC-wPBE(w=0.4) at a range of basis sets with geometry optimization at the same method and basis.

Figure 7 shows p delta plus s delta for formaldehyde, formamide and N-methylformamide, revealing the enhancing role of amide resonance in these HB.

Ap1:Figure 32 shows selected steric interactions to be considered in explaining why strongest HB does not occur at 90 degrees. Two of these rise sharply between 75 and 90 degrees. F(lp-p-2)|C-H is 0.6 kcal/mol at 75 degrees and 1.72 kcal/mol at 90 degrees. H-F|C-H is 0.23 kcal/mol at 75 degrees and 1.10 kcal/mol at 90 degrees. Together, these interactions offer explanation that it is steric interaction between HF and the hydrogen of the carbonyl carbon that prevent strongest hydrogen bonding occurring closer 90 degrees.

Ap1:Figure 33 shows H-F* NBO occupancy as C-O-F angle is varied from 45 to -45 degrees, revealing imperfect symmetry about 0 which is presumably due to interactions between HF and the methyl group attached to the amide N and the difference in electrostatic field caused by HF.

Ap1:Figure 34, Ap1:Figure 35, Ap1:Figure 36 and Ap1:Figure 37 show the results of unconstrained geometry optimization from a range of initial C-O-F angles for different methods, basis sets and with dispersion corrections. At least at the 1 degree gradations used for initial angle, there is no initial angle close to 0 where the optimized angle is also close to zero. If a range of initial angles exists that balances the tendency of HF to tip one way versus the other exists, this range must be less than 1 degree. In being free from constraint, this set of experiments considers all factors, including electrostatic interactions and dispersion such as captured by respective methods, and is consistent with experiments that focus on donor-acceptor and steric interactions alone. Neither electrostatic interactions nor dispersion at the applied correction result in linear or near linear C-O-F geometry being preferred.

In Figure 8 and Ap1:Figure 38, the C-O-F angle is fixed and the rotation is away from the amide plane. Ap1:Figure 38 shows H-F* NBO occupancy at a wavefunction method, and Figure 8 shows p delta plus s delta, necessarily at a non-correlated method, and is more revealing of the relationship between each given C-O-F angle. Moving out of the amide plane reduces p delta plus s delta due to reduced interaction with the p-type lone pair.

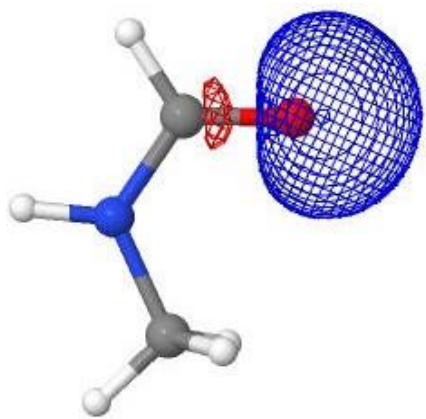

Figure 1. s-rich Amide Oxygen Lone Pair NBO

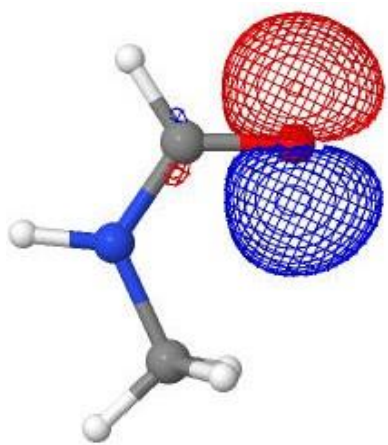

Figure 2. p-type Amide Oxygen Lone Pair NBO

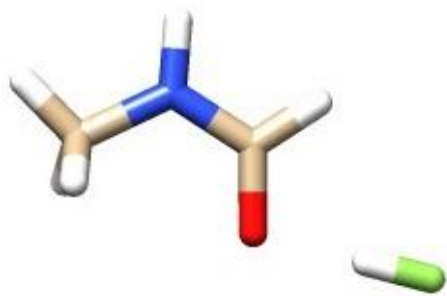

Figure 3. HF Hydrogen Bonded to N-methylformamide O at C-O-F Divergence from Linear of 75 degrees in Amide Plane with F Distal to N

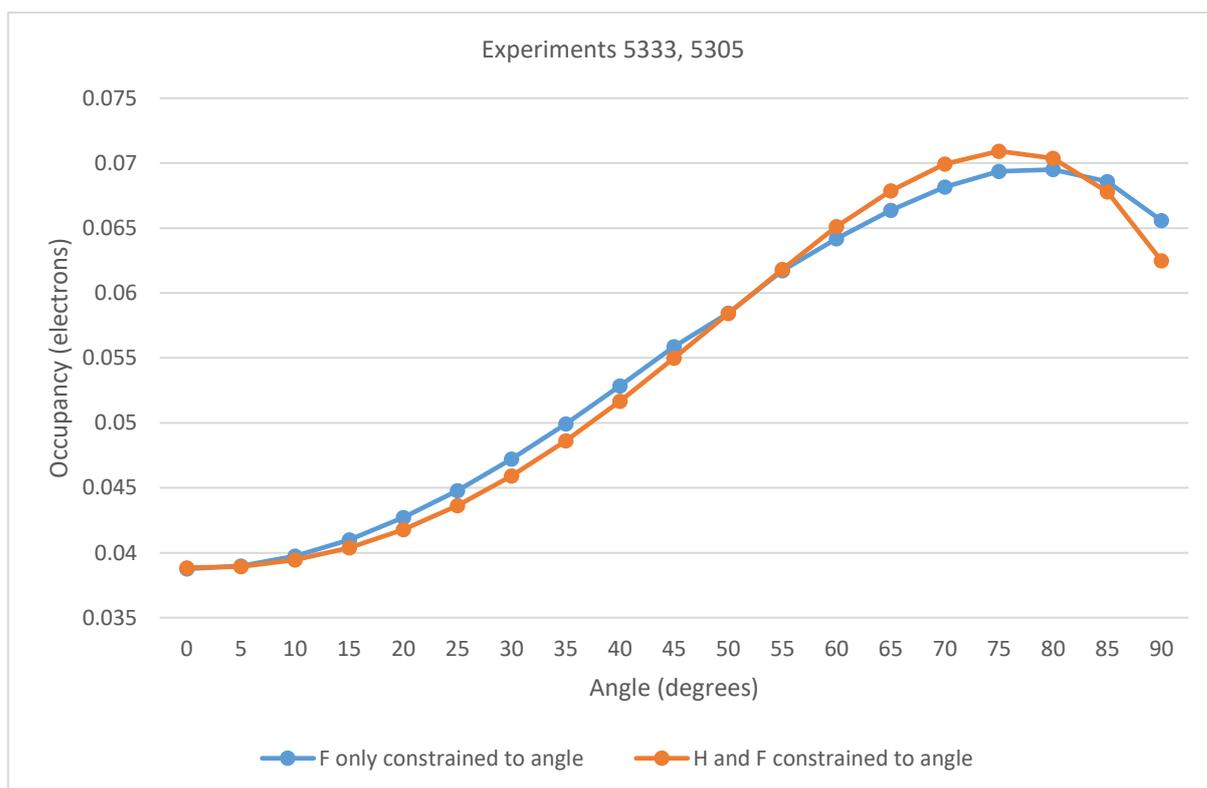

Figure 4. H-F* NBO Occupancy with F or H and F Constrained to Angle from C-O at O of N-methylformamide with HF Distal to N and in Amide Plane at SCS-MP2/aug-cc-pVTZ

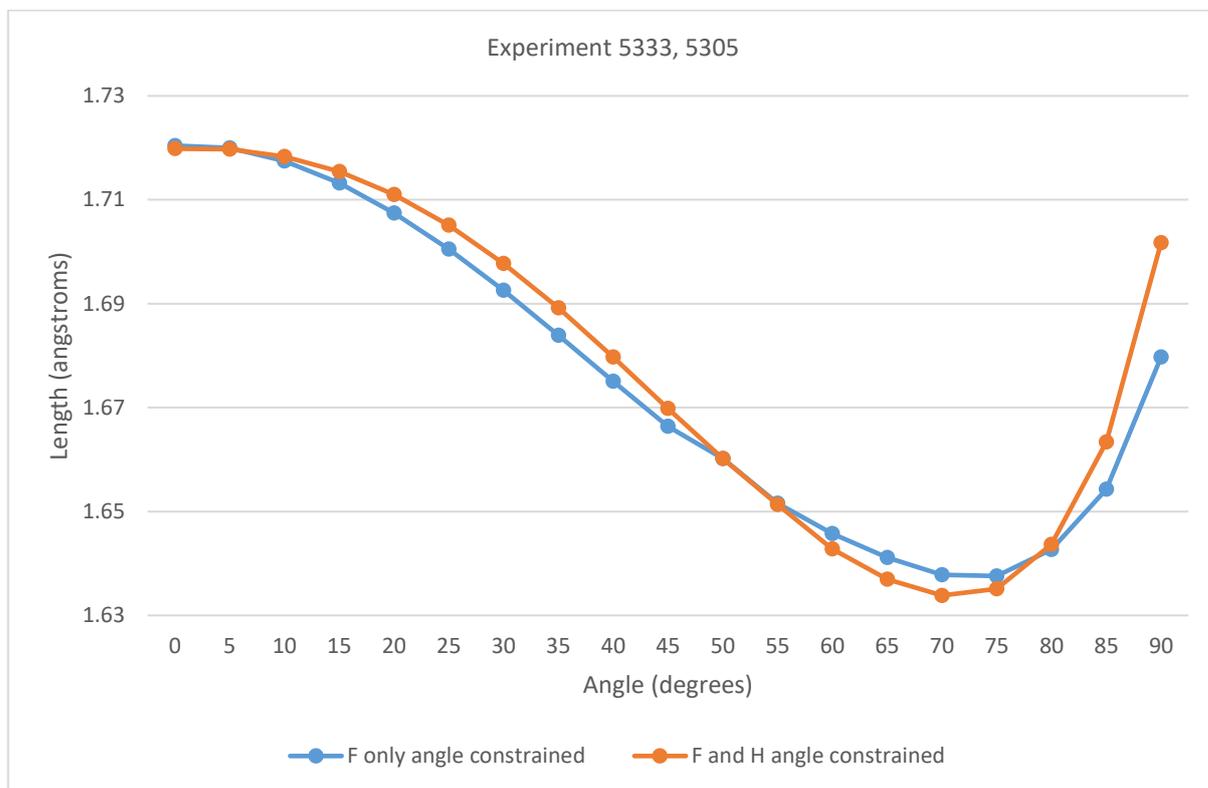

Figure 5. N-methylformamide/HF Hydrogen Bond Length with F Only or F and H Constrained to Angle From C-O at O with HF Distal to N and in Amide Plane at SCVS-MP2/aug-cc-pVTZ

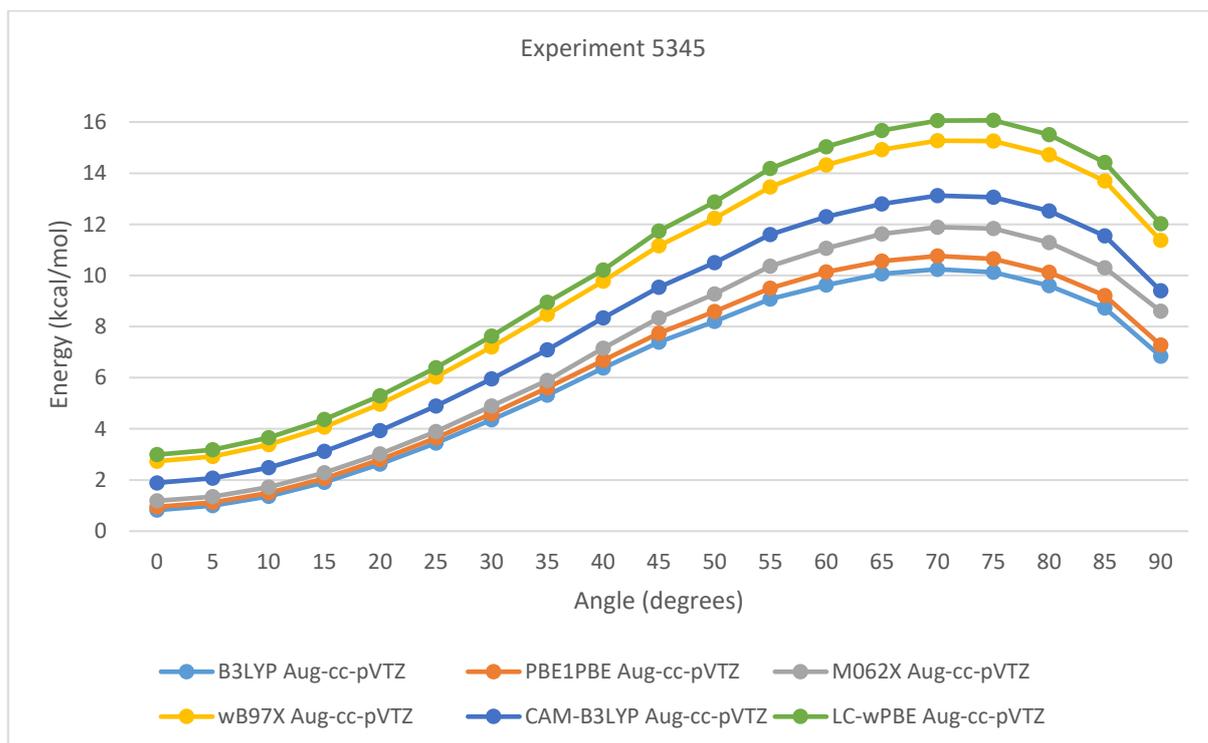

Figure 6. Donor-Acceptor SOPT Energy Minus Steric Exchange Energy for Interactions Between N-methylformamide O Lone Pairs and H-F and H-F* at C-O-F Angle in Amide Plane with Geometry Optimized at SCS-MP2/aug-cc-pVTZ

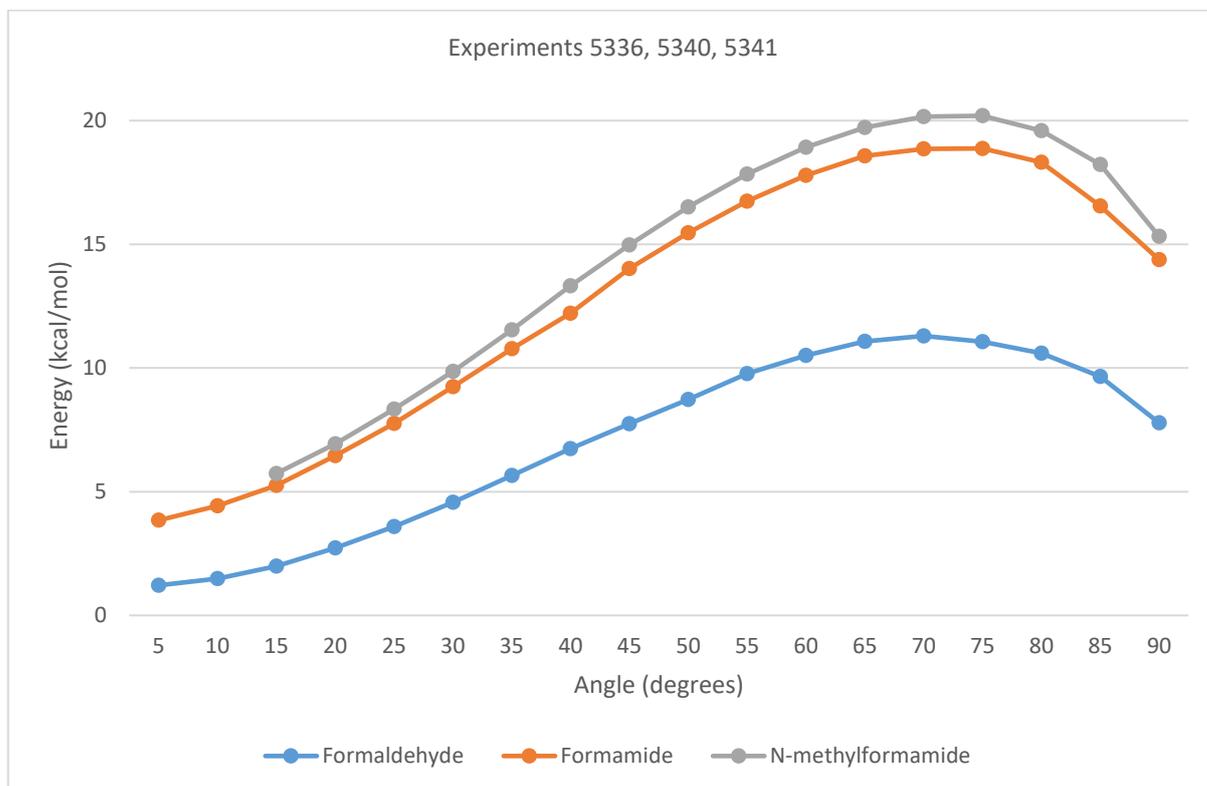

Figure 7. Donor-Acceptor SOPT Energy Minus Steric Exchange Energy for Interactions of O Lone Pairs and H-F and H-F* at C-O-F Angle in Aldehyde or Amide Plane at LC-wPBE(w=0.4)/aug-cc-pVTZ

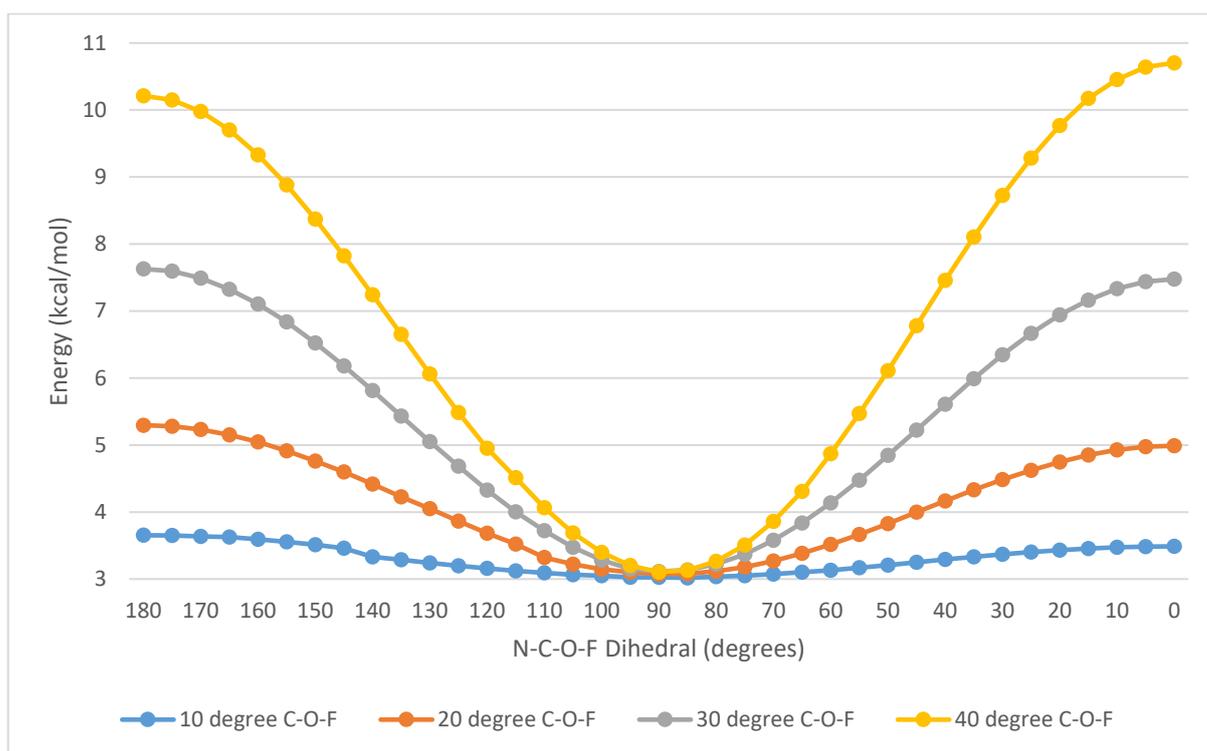

Figure 8. Donor-Acceptor SOPT Energies minus Steric Exchange Energies for O Lone Pair Interactions with H-F and H-F* with HF Hydrogen Bonded to N-methylformamide O at Constant C-O-F Angle with F Rotated about C-O Axis at LC-wPBE(w=0.4)/aug-cc-pVTZ over SCS-MP2/aug-cc-pVTZ Optimized Coordinates

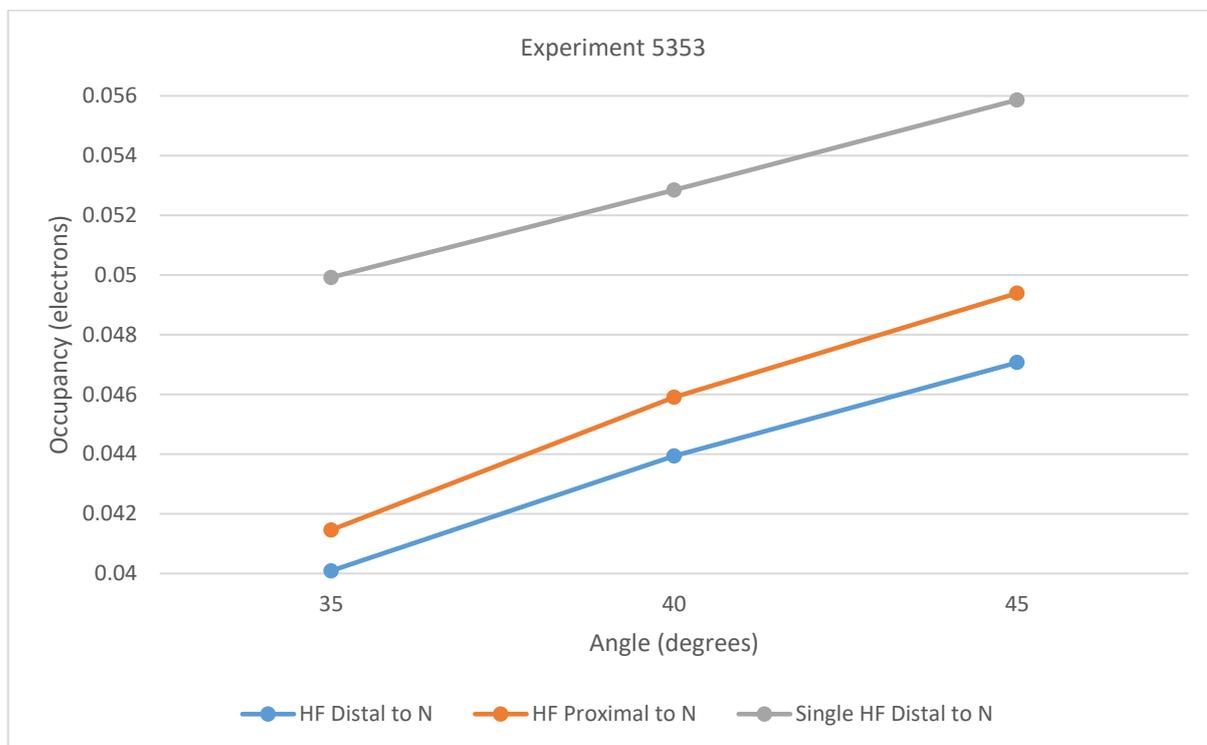

Figure 9. H-F* NBO Occupancies with 2 HF Hydrogen Bonded to N-methylformamide O with Both HF in Amide Plane at Same C-O-F Angle Each Side of C-O Compared with Single HF at SCS-MP2/aug-cc-pVTZ

### 5.1.2 Use of hydrogen bonding to carbonyl oxygen where C-O-N is far from linear

Figure 9 shows H-F* NBO occupancy for the case of two HF molecules in the amide plane, one distal to N, one proximal to N, both at the same C-O-F angle. The range of C-O-F is restricted to minimize the interactions between the HF molecules and the interactions of one of them with the N methyl group. The H-F* occupancy of the distal HF is about 25% less than that of the corresponding single distal HF case, and the proximal HF case has about 22% less occupancy of the single distal case. Reduction due to the busy donor effect [1] is to be expected, but it is noteworthy that the sum of the occupancies of the two HF case is considerably more than the occupancy of the single HF case. The HF interactions in the 2 HF case are anti-cooperative in keeping with the busy donor effect, but the total charge transfer and hence resonance in moieties mediating RAHB suggests an opportunity to develop new HB polymers with greater stability than polyamides [45] and polyurethanes [46] by improving on the RAHB of polymers through a bifurcation geometry more favourable than that of urea-based polymers [47]. In urea-based polymers, two H-N bonds parallel to the C-O of the next chain, giving bifurcated HB at O but at distinctly sub-optimal geometry (Ap1:Figure 39), whereas optimal geometry is to be found by the H-N bonds pointing in to the O in the manner of 2 HF HB to amide oxygen experiment above (Figure 9). This bifurcation geometry improvement requires that the nitrogens be further apart than they are in urea. One HB to urea oxygen at large C-O..N angle from linear will transfer substantially more charge to H-N* than an HB at shallower angle from linear, and two at large angle will deliver substantially more in total again though less than double. The total resonance of the resonant moieties connecting the HBs will be markedly increased in the progression through the three cases: one HB with linear C-O-N, one HB with large C-O-N divergence from linear, two HBs with large C-O-N divergence from linear. The design problem is to have near-optimal bifurcation geometry and still have a highly resonant moiety present, and a question is at what atom count such a solution might be found should such solutions exist. Evolutionary algorithms [48] might be used to search the space of possibilities.

Large C-O-N divergence from linearity by non-bifurcated hydrogen bonds is seen in nature. There are examples of this in nitrogenous base pairing [49] such as in guanine/cytosine pairing. HB of the carboxyl group to guanidinium demonstrates C-O-N at ~63.5 degrees from linear (Figure 40). Replacing formate with carbamate (Figure 41) gives yet closer hydrogen bonding, interpretable as being due to 2 charge transfers from carbamate N to its carboxyl group. When guanidinium is replaced with urea (Figure 42 and Figure 43), bond lengths increase due to urea not having the positive charge of guanidinium to neutralize the charge of the carboxyl group. The overall charge of a unit of the polymer must be zero. These examples of large C-O-N divergence for a single HB might serve as a starting point for design of optimally bifurcated HB at double bonded oxygen or sulfur.

Smart rubbers [50] make no use of covalent chains for their assembly, and entirely rely on HB for their properties. If the design of smart rubber emphasized the HB geometries discussed here, strength of the material might be improved such that the strength of ordinary rubber might be reached.

Much interest is now focussed on development of materials based on covalently bonded sheets and cylinders for nanomaterials applications [51], but desirable properties might still be found with HB polymers, particularly in view of the substantial increase in total resonance and hence stabilization potentially available with more favourable HB bifurcation geometry at double bonded oxygen or sulfur.

### 5.1.3 Experiments involving hydrogen bonding between two amide groups

A pair of N-methylformamides is used to investigate hydrogen bonding between a pair of amide groups. Figure 10 shows the sum of the inter amide donor-acceptor SOPT energies minus the sum of the inter-amide steric exchange energies. It is remarkable how low the total is, particularly with the aug-cc-pVTZ basis set. As the angle approaches 0, for some method/basis combinations the result become negative, but otherwise they are quite modest positive figures. That inter-amide donor-acceptor minus steric energetics for HB are minor between 2 co-planar amides and even negative close to C-O-H linearity depending on method/basis used is a surprising result. Ap1:Figure 44 separates the SOPT and steric exchange figures for the inter-amide interactions for 1 method and 2 basis sets. Ap1:Figure 45 shows select SOPT and steric energies internal to one amide and Ap1:Figure 46 internal to the other. Ap1:Figure 47 shows the N(lp)->C-O(p)* SOPT energy minus steric exchange energy for each amide with 1 method and 2 basis sets.

Ap1:Table 6 gives these internal amide figures for one N-methylformamide in isolation. The N(lp)|C-O(p) steric exchange differs little between this case or either of the bound amides, but the N(lp)->C-O(p)* is ~112 kcal/mol in each of the bound amides at 0 degrees, but in the isolated amide is about 102 kcal/mol, so that the hydrogen bonding at 0 degrees is about 20 kcal/mol due to amide resonance alone. It is not charge transfer minus steric interactions between amides that predominate in hydrogen bonding between amides, it is the change in the resonance of each amide. The interactions between the amides serve to increase the amide resonance in each amide, rather than having direct bonding energetic significance of their own. This is at odds with any view that the direct inter-amide interactions are energetically dominant with the increase in resonance of each amide relegated to the role of bonus. That the majority of bonding energetics reside in the increase of the amide resonance has particular significance in terms of the findings of [4]. Inter-amide hydrogen bonding is largely rather than partially susceptible to variation of amide resonance by electrostatic field with component parallel to the amide C-N bond, increasing the likely significance of this variation relative to other factors in protein folding such as pure electrostatics, hydrophobia and entropy.

**Notes for Figure 10**

Amide 1 has the oxygen participating in the HB, Amide 2 the hydrogen of the same HB

The donor-acceptor (SOPT) interactions between the amides summed are:

Amide 1 -> Amide 2

O(lp-p) -> H-N*, C-O(p)*, C-O(s)*

O(lp-s) -> H-N*, C-O(p)*, C-O(s)*

C-O(p) -> H-N*

C-O(s) -> H-N*

Amide 2 -> Amide 1

H-N -> C-O(p)*

H-N -> C-O(s)*

The steric interactions between the amides summed are:

Amide 2 | Amide 1

N(lp) | C-O(p), C-O(s), O(lp-p), O(lp-s)

H-N | C-O(p), C-O(s), O(lp-p), O(lp-s)

C-O(p) | O(lp-p), O(lp-s)

C-O(s) | O(lp-p), O(lp-s)

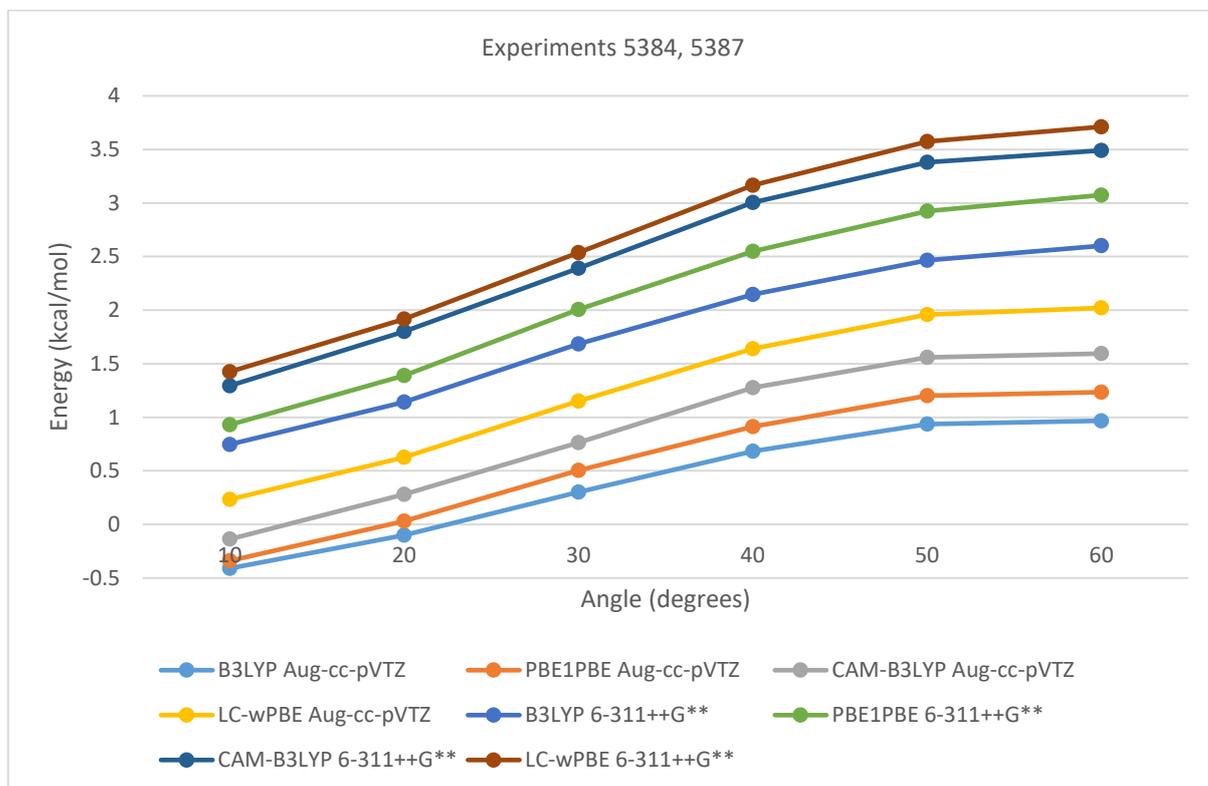

Figure 10. Total SOPT Energy Minus Steric Exchange Energy Between 2 Coplanar Hydrogen Bonded N-methylformamides at Given C1-O1-N2 Angles with N1-C1-O1-N2 Dihedral 180 Degrees

## 5.2 Inter-backbone amide hydrogen bonding C-O-N angles in the Protein Data Bank

The Protein Data Bank was queried by the means described in [4] for backbone amide-backbone amide hydrogen bonding with geometry restricted as defined in the notes for the resulting Ap1:Table 7. This data suggests that C-O-N angles with the large non-linearity investigated above are sterically disfavoured in proteins. The method for querying the PDB is further described in Section 5.7 (Cyclic HB in the Protein Data Bank).

## 5.3 Linear chain of formamides

A linear chain of 8 coplanar formamides was geometry optimized at LC-wPBE(w=0.4)/6-311++G** without constraint. The chain remained coplanar (Figure 11). Figure 12 shows that the SOPT energy associated with the primary amide charge transfers follows the expected pattern, peaking in the middle of the chain. Ap1:Figure 48 shows that interactions between the formamide units that may be responsible for maintaining approximate linearity of the chain remain minor, though O(lp-s)->H-C* is most notable. Figure 49 shows that O(lp-p)->H-N* exceeds O(lp-s)->H-N* by the second hydrogen bond, and the latter declines more than the former in the final hydrogen bond which is in keeping with the increased C-O-N (less linear) angle shown in Figure 13. The resonance of the preceding amides is related to the C-O-N angle, except where the resonance has increased O(lp-s)->H-C*.

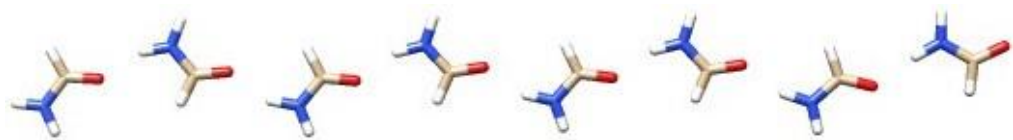

Figure 11. Chain of 8 Formamides after Unconstrained Geometry Optimization at LC-wPBE(w=0.4)/6-311++G**

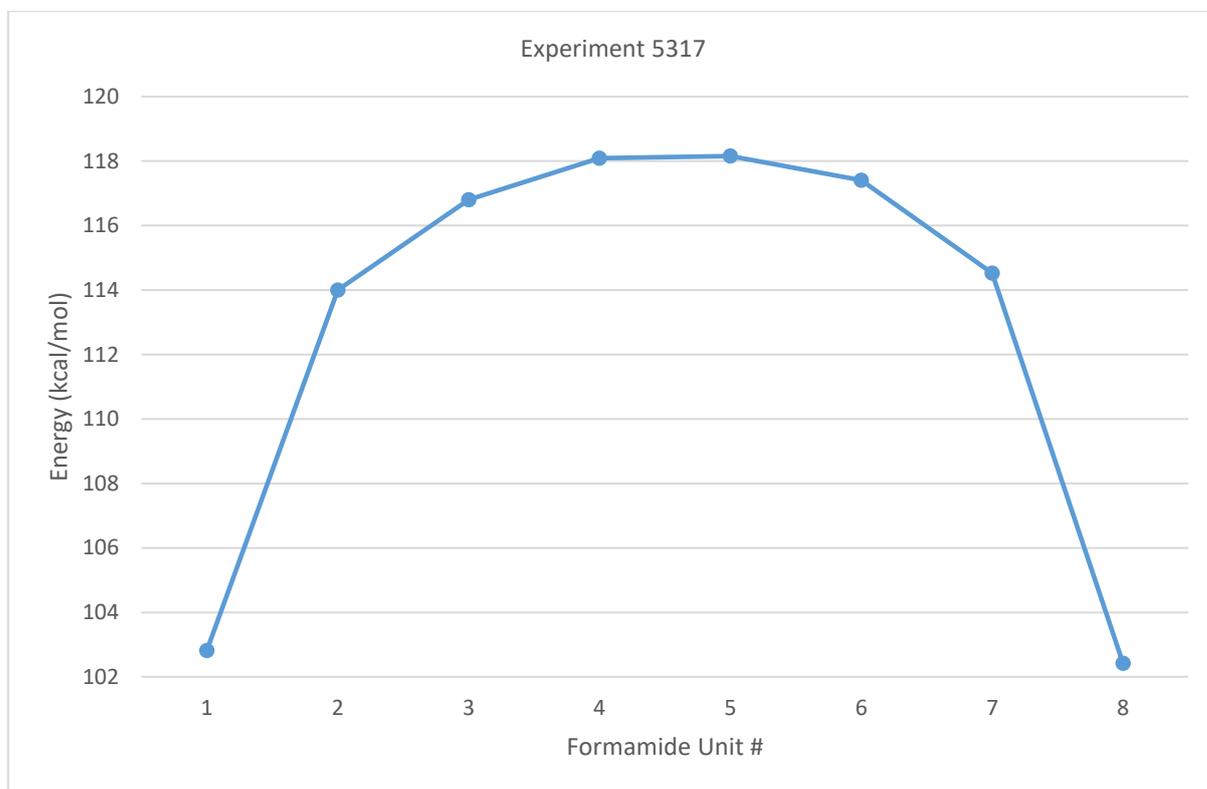

Figure 12. N(lp)->C-O(p)* SOPT Energy in Hydrogen Bonded Chain of 8 Formamide Units with Unconstrained Geometry Optimization at LC-wPBE(w=0.4)/6-311++G**

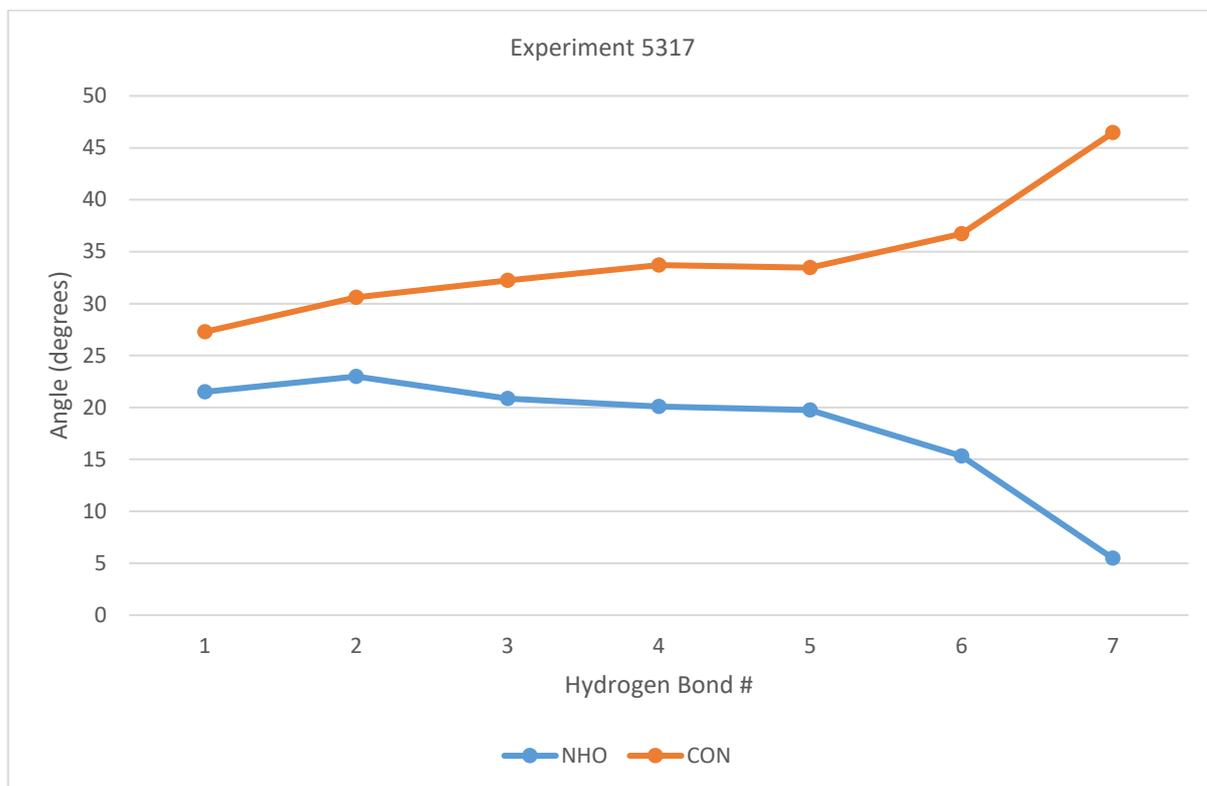

Figure 13. Hydrogen Bond Angles in Chain of 8 Formamide Units with Unconstrained Geometry Optimization at LC-wPBE(w=0.4)/6-311++G**

### 5.4 Formamide cycles

Cycles from 6 to 12 coplanar units of formamide such as in Figure 14 were geometry optimized without constraint at LC-wPBE(w=0.4)/6-311++G**, and these remained coplanar. It can be seen in Figure 15 that hydrogen bond length is a minimum at 12 units, though is not uniformly declining from 6 units due to an increase at 11. In Figure 16 it can be seen that the primary amide charge transfer energy peaks at 8 units and has a downward spike at 11 units. This is in keeping with the SOPT minus steric exchange energy line seen in Ap1:Figure 50, though the slope of the line is shallow. As seen in Figure 17, the maximum variation in C-O-N angle with a cycle is at 11 units and the minimum is at 8 units. Also, the minimum variation in N-H-O angle within a cycle occurs at 8 units.

There are two limitations to cyclic hydrogen bonding seen in this section. The first is that when coplanar, the average C-O-N angle necessarily declines with increase in the number of units in the cycle, and eventually amide resonance declines, and with it hydrogen bonding. Coplanarity does not favour large cooperative cycles. The second is the sensitivity to uniformity of C-O-N angle in a cycle, since the least amide resonance in the cycle is limiting of cyclic RAHB.

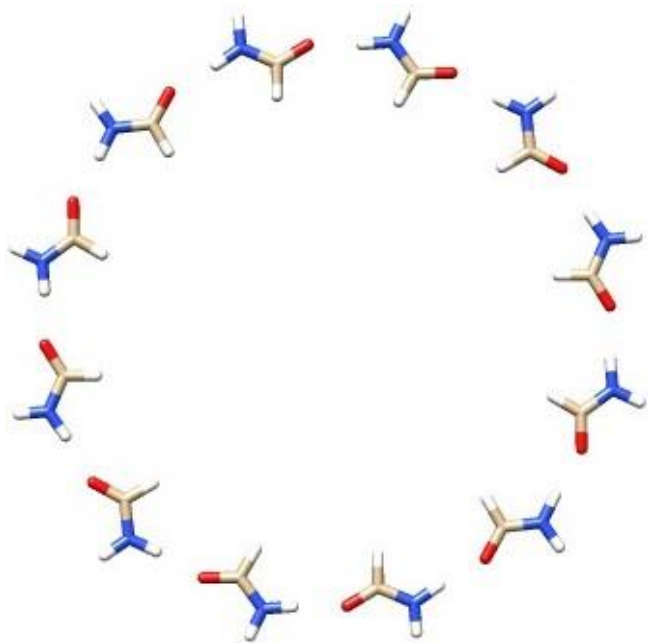

Figure 14. Cycle of 12 Formamides after Unconstrained Geometry Optimization at LC-wPBE(w=0.4)/6-311++G**

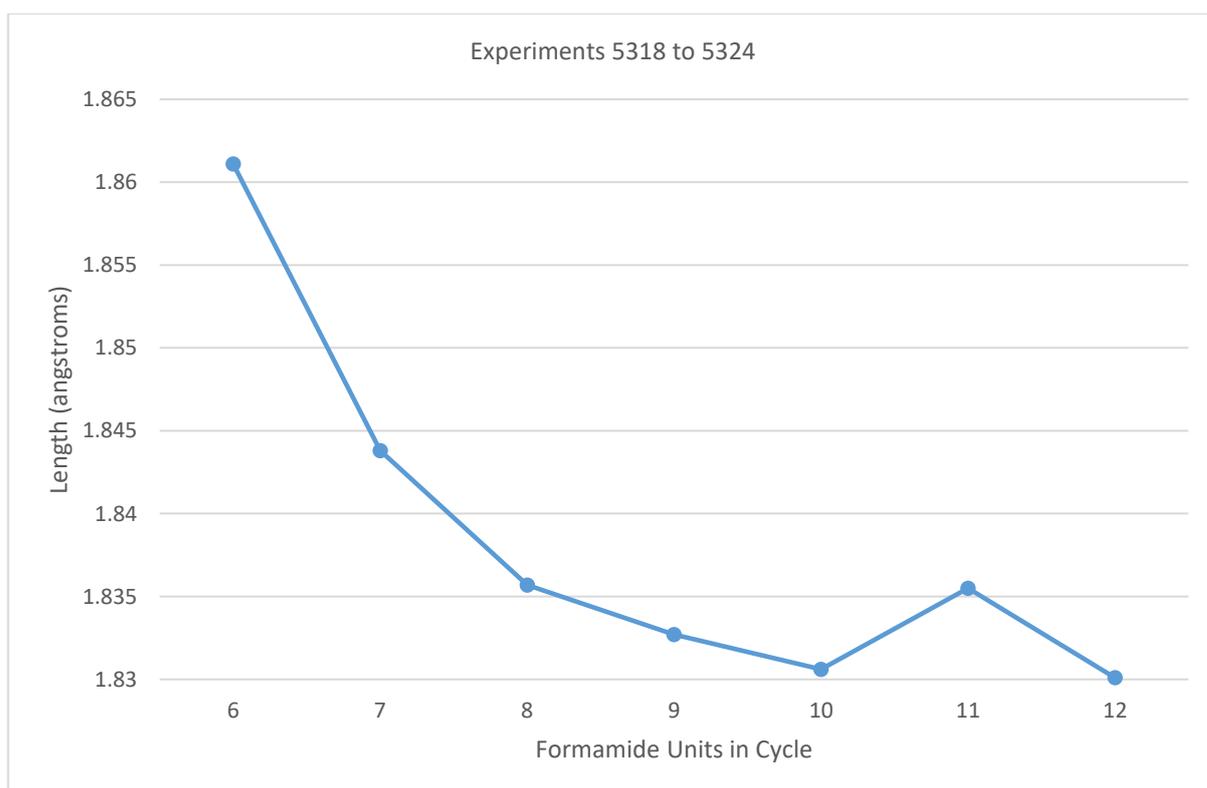

Figure 15. Hydrogen Bond Length in Planar Cycles of Formamide Optimized at LC-wPBE(w=0.4)/6-311++G**

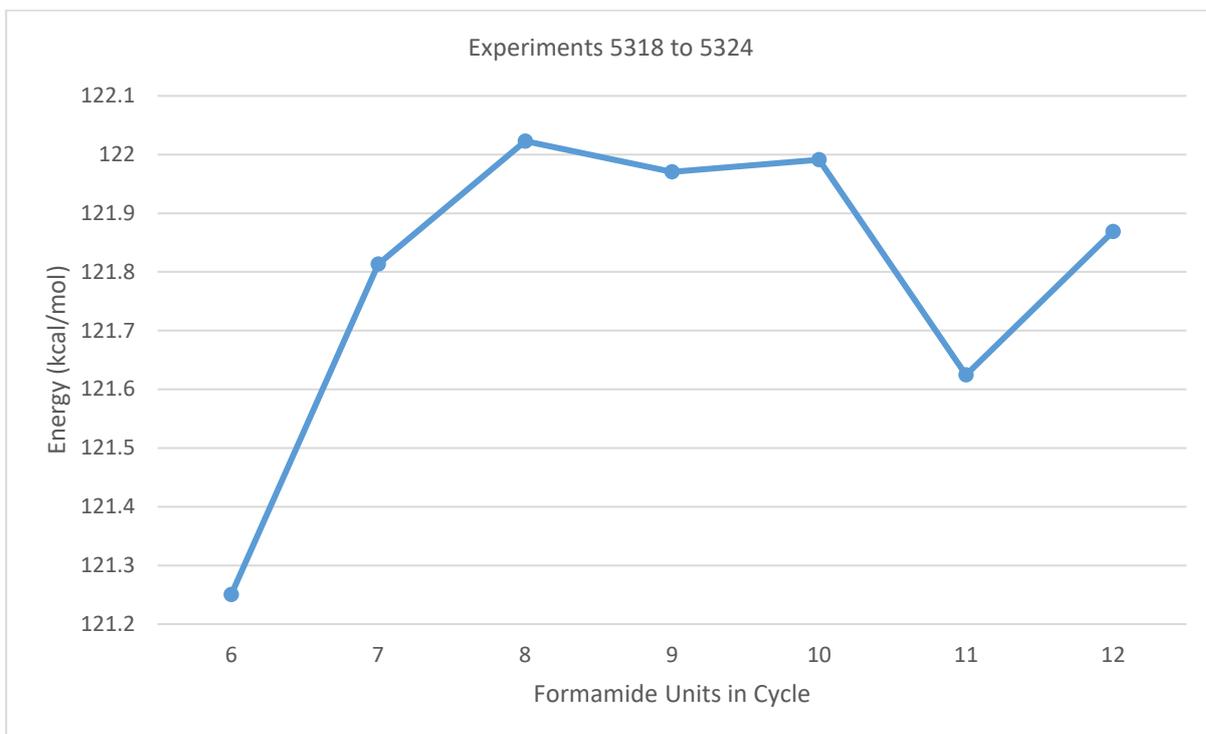

Figure 16. N(lp)->C-O(p)* SOPT in Hydrogen Bonded Formamide Units in Planar Cycle Optimized at LC-wPBE(w=0.4)/6-311++G**

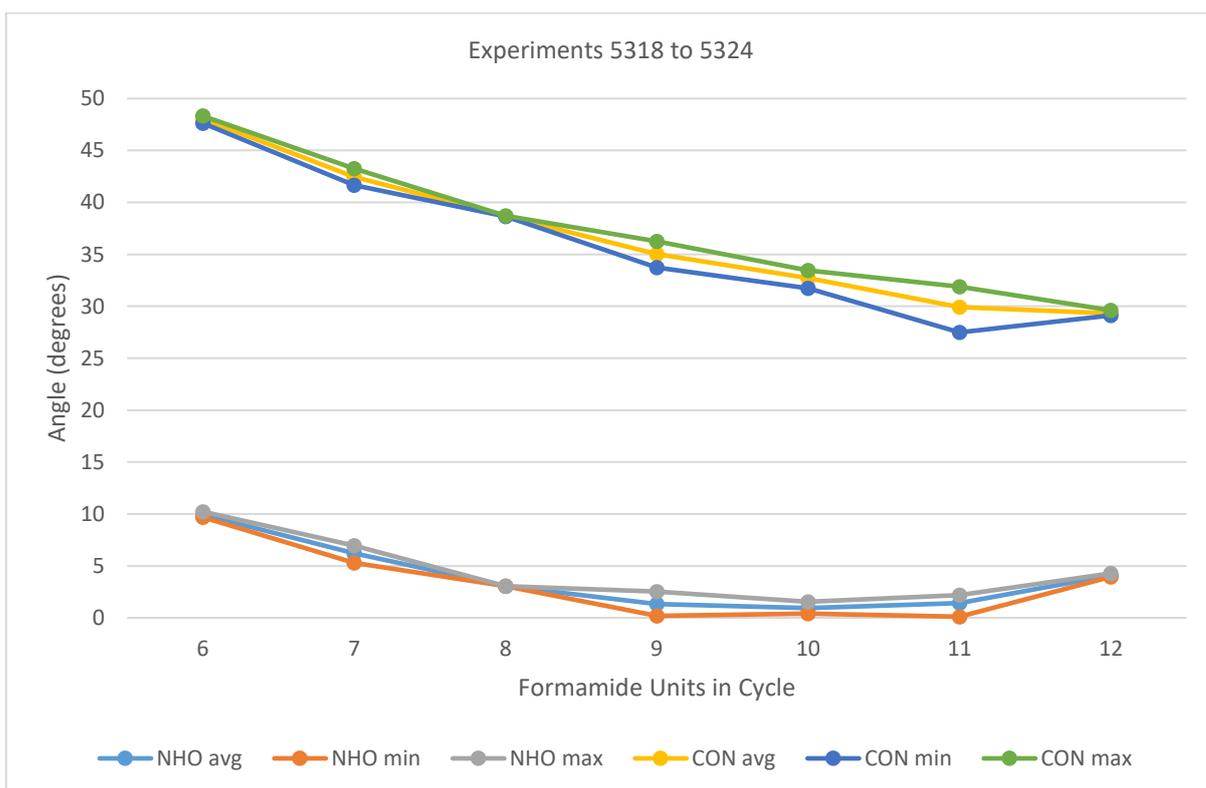

Figure 17. Hydrogen Bond Angles in Planar Cycles of Formamide Optimized at LC-wPBE(w=0.4)/6-311++G**

## 5.5 Three alpha helices

In exploration of what binding energy advantage exists in non-dynamic gas-phase cycles of hydrogen bonding, three Ace-Ala(15)-Nme alpha helices were connected by three formamides such that each formamide hydrogen bonds with two spines of each of two helices (Figure 18). Only one spine in each helix is hydrogen bonded by two formamides, and only this spine participates in a hydrogen bonding cycle (Figure 19). Formamide is representative of the sidegroup of asparagine or glutamine. HB and amide resonance of this three helix arrangement is compared with those of a single such helix with two capping formamides (Figure 20) in the manner of the three helix arrangement and the result is shown in Table 1.

The key to evaluating cyclic RAHB is the primary amide resonance charge transfers, which increase from ~1626 to ~1693 kcal/mol for backbone amides per helix, and decrease from ~108 to ~96 kcal/mol for each formamide. The cyclization is thus favourable by ~201 kcal/mol for the three-helix arrangement as assessed by primary backbone amide resonance charge transfer alone. In a physiological situation, the HB sites of the formamides, representing sidechain amides, in the single helix case are likely to be fully utilized, which would increase the resonance of the amides of the helix, but these isolated, gas-phase and non-dynamic experiments demonstrate that cyclization is advantageous under isolated circumstances.

The "ressb" column, N(lp)->C-O(s)*, of Table 1 shows purely DFT error and indicates that the findings of [11] for beta sheets extend to alpha helices although they are less marked. The "ressb" column shows this DFT error for formamide, indicative of the error in asparagine and glutamine.

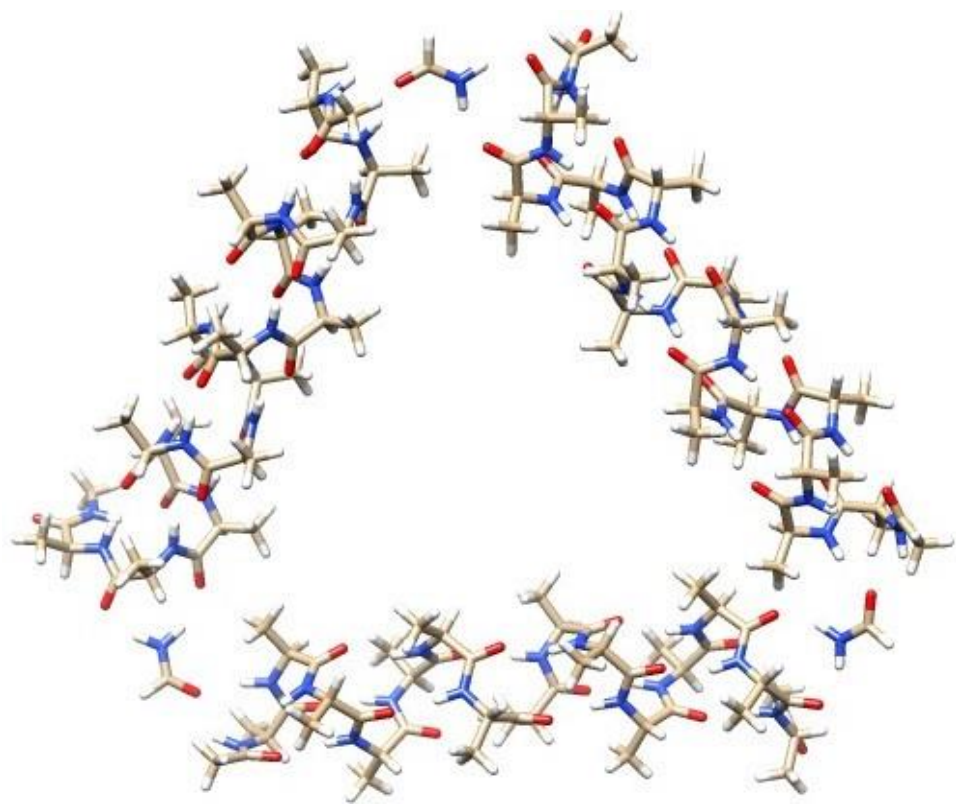

Figure 18. 3 ACE-Ala(15)-NME Alpha Helices with 3 Formamides Each Hydrogen Bond-Connecting Two Spines of Adjacent Helices at LC-wPBE(w=0.4)/6-311++G**

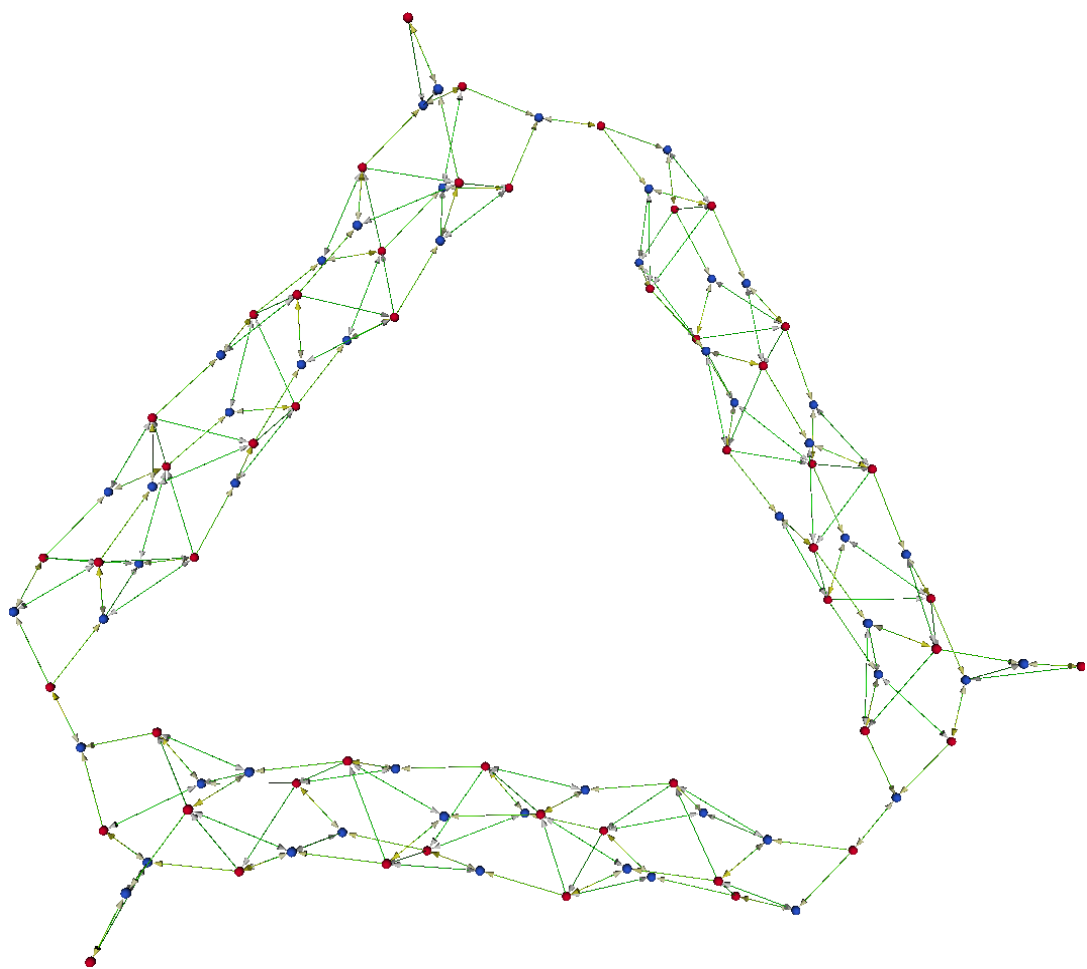

Figure 19. Representation of Primary Amide Charge Transfers and Inter-Amide Hydrogen Bond Charge Transfers to Show Hydrogen Bonded Chains of Amides of Figure 18. Red Spheres Amide O, Blue Spheres Amide N. Primary Amide CT Shown as From N to O, Inter-Amide CT as O to N

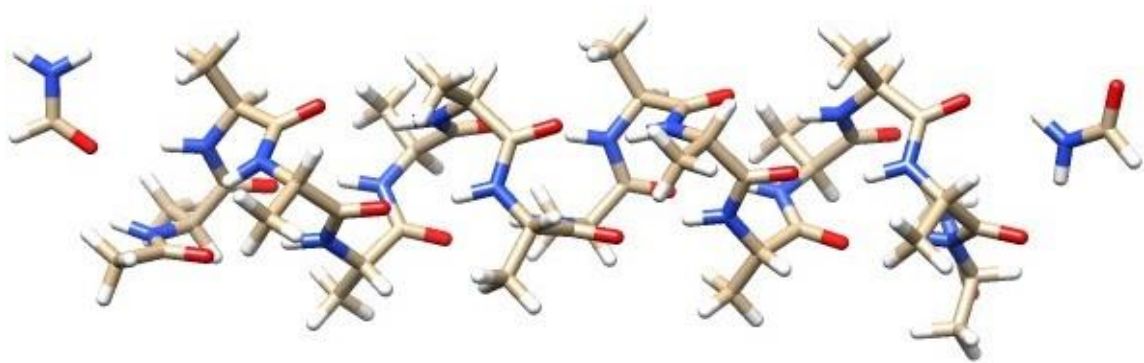

Figure 20. ACE-Ala(15)-NME Alpha Helix with a Formamide Hydrogen Bonded at Each Terminus at LC-wPBE(w=0.4)/6-311++G**

Table 1. Donor-Acceptor Interactions for 1 Alpha Helix versus 3 Alpha Helices with some Spines Connected by Formamides

| helices | hbp | hbs | hbpr | hbsr | hbpl | hbsl | respb | ressb | resps | resss |
|---|---|---|---|---|---|---|---|---|---|---|
| 1 | 22.6 | 67.87 | 4.88 | 5.57 | 4.69 | 6.27 | 1625.62 | 24.89 | 108.02 | 0.21 |
| 3 | 28.01 | 90.93 | 10.79 | 13.03 | 8.77 | 12.18 | 1693.11 | 24.92 | 95.93 | 15.53 |

- hbp: O(lp-p)->H-N* kcal/mol average of intra helix totals
- hbs: O(lp-s)->H-N* kcal/mol average of intra helix totals
- hbpr: O(lp-p)->H-N* kcal/mol average of helix to formamide HB CT totals
- hbsr: O(lp-s)->H-N* kcal/mol average of helix to formamide HB CT totals
- hbpl: O(lp-p)->H-N* kcal/mol average of formamide to helix HB CT totals
- hbsl: O(lp-s)->H-N* kcal/mol average of formamide to helix HB CT totals
- respb: amide resonance N(lp)->C-O(pi)* kcal/mol average of helix totals
- ressb: amide resonance N(lp)->C-O(sigma)* kcal/mol average of helix totals
- resps: amide resonance N(lp)->C-O(pi)* kcal/mol average of formamide totals
- resss: amide resonance N(lp)->C-O(sigma)* kcal/mol average of formamide totals

The 1 helix case is shown in Figure 20 and the 3 helix case in Figure 18.

## 5.6 Artificial beta sheets

Appendix 2 (Section 10) provides tabulation of polyvaline parallel beta sheet backbone amide N(lp)->C-O(p)* and N(lp)->C-O(s)* SOPT energy values for 4 backbone HB chain, 4 beta strand beta sheets, capped with formamides, some configurations forming HB cycles. The particular configuration for each experiment is shown in the image prefacing each experiment. The image is laid out so that the position of the backbone amides correspond to the positions of the cells in the tables for the experiment.

Appendix 3 (Section 11) provides the same manner of tabulation for polyvaline antiparallel beta sheet capping, and includes some examples of water capping the beta sheets though this study is focussed on cycles of protein.

In these experiments, each beta sheet including capping molecules was geometry optimized without constraint at LC-wPBE(w=0.4)/6-311G**. TeraChem was used for the geometry optimization (its method name for LC-wPBE is wpbe) and Gaussian for the production of the NBO 6 GenNBO input .47 file.

As mentioned above, non-zero N(lp)->C-O(s)* values are purely DFT error. While the total of this value is large for each experiment (Ap2:Table 18 for parallel sheet and Ap3:Table 29 for antiparallel sheet), it is the distribution of these errors that is most alarming. Inspection of the 10 N(lp)->C-O(s)* tables in Appendices 2 and 3 gives that notable instances of this error are sparse rather than evenly distributed.

This is alarming because a N(lp)->C-O(p)* value in an RAHB chain is partially determined by the other such values in the chain, even where such a value is not associated at the same amide with a significant N(lp)->C-O(s)* value. It is clear that a large reduction in N(lp)->C-O(p)* value is associated with a large N(lp)->C-O(s)* value at an amide, so one large N(lp)->C-O(s)* in an RAHB chain invalidates all N(lp)->C-O(p)* values in the chain. 6-311G** can be taken to be unsatisfactory for use with beta sheets. This error should be taken as influencing geometry optimization as well as single point orbitals. In the presence of such errors, we do not proceed to discuss the tabulated values for cyclic hydrogen bonding as relevant to the study of nature. The study of RAHB cyclization in beta sheets awaits methods that do not suffer the errors described in [11] and otherwise give a good account of RAHB and computationally scale to the atom count of beta sheets.

## 5.7 Cyclic HB in the Protein Data Bank

A subset defined below of the NMR entries of the PDB was searched for cyclic HB in proteins preparatory to investigating any cooperativity in these cycles with quantum chemical methods. Non-standard residues, ligands and solvent were not considered. Data derived from NMR spectroscopy but not X-ray crystallography was used, for the protonation state of residues was required and the observed coordinates of these protons desirable. The restriction of the search to NMR spectroscopy-derived data biases the search toward smaller proteins.

Only the first model in a PDB file was considered. Also, PDB files were excluded if any error in the file was detected. The programs for spatial query of the PDB were intended to support a broad range of queries that are as straightforward as possible, and the principle that queries should not have to handle error or exception conditions was adopted, so files with these conditions were excluded from querying. Conditions evaluated were: values expected to be integer actually integer, residue sequence numbers in a chain contiguous, no atoms missing, helix start residue sequence number less than helix end residue number, helix type in the range 1..10, start and end residues of helices and beta strands being of the same chain, no secondary structures overlapping and the backbone nitrogen at the N-terminus of a chain having 3 hydrogens.

Each potentially cooperative unit was modelled as a list of polar hydrogens and a list of HB acceptor (charge transfer donor) atoms, with the connection between hydrogen and acceptor atoms within the cooperative unit left abstract until later selection for quantum chemical analysis. Determination of the extent of cooperativity was similarly deferred, and all protein HB cycles regardless of cooperativity were captured in the PDB extraction pass.

Modelling of backbone amides and sidechain amides as cooperative units was straightforward. Aspartate and glutamate sidechains have only acceptors unless a proton is bound which then allows two cooperative functional units, one being the new hydroxyl, the other being a larger unit H-O-C-O which somewhat resembles the amide H-N-C-O. Singly protonated histidine has both an acceptor and a polar hydrogen. The hydroxyl groups of serine, threonine and tyrosine were modelled as cooperative units.

The maximum HB length is taken to be 3.0 angstroms and X-H..A angle of 45 degrees as an HB in a significantly cooperative system will have length and angle appreciably less than these figures which then form an upper bound.

8378 PDB files were extracted from the RCSB Protein Data Bank on 2014-9-8 by Advanced Search with the following parameters

- Macromolecule
    - Contains Protein = Yes
    - Contains DNA = No
    - Contains RNA = No
    - Contains DNA/RNA Hybrid = No
- Experimental Method
    - Solution NMR
    - Has Experimental Data = Ignore
- Has Modified Residue(s) = No

This can be closely reproduced by further specifying a Release Date of up to 2014-9-8, returning 8377 files, the discrepancy of 1 perhaps attributable to a new file being made available during that day.

1690 files of these 8378 PDB files were filtered out as having error or exception conditions and the first model of the remaining 6,688 files was returned for querying. These were programmatically analysed for cycles of HB, and the resulting data appears as Table 2.

Table 2. Cycles of hydrogen bonding in proteins

| PDB | U1 BA/SC | Res1 | Seq1 | Ch1 | Struct1 | U2 BA/SC | Res2 | Seq2 | Ch2 | Struct2 | U3 BA/SC | Res3 | Seq3 | Ch3 | Struct3 | U4 BA/SC | Res4 | Seq4 | Ch4 | Struct4 |
|---|---|---|---|---|---|---|---|---|---|---|---|---|---|---|---|---|---|---|---|---|
| 2U2F | BA | GLY | 7 | A | Strand=(1,S3) | BA | VAL | 75 | A | Strand=(1,S4) | BA | LEU | 74 | A | Strand=(1,S4) | BA | GLY | 8 | A | Coil |
| 2M89 | SC | ASN | 115 | B | Helix=(RH Alpha,8) | SC | ASN | 115 | A | Helix=(RH Alpha,4) | | | | | | | | | | |
| 2LKD | SC | ASN | 53 | A | Coil | BA | ILE | 57 | A | Coil | BA | LYS | 56 | A | Coil | | | | | |
| 2KQ2 | BA | ASP | 12 | A | Coil | SC | ASN | 124 | A | Helix=(RH Alpha,7) | BA | GLY | 13 | A | Coil | | | | | |
| 2KHQ | SC | GLN | 53 | A | Helix=(RH Alpha,5) | SC | ASN | 57 | A | Helix=(RH Alpha,5) | | | | | | | | | | |
| 2K2W | SC | ASN | 262 | A | Helix=(RH Alpha,3) | SC | THR | 265 | A | Strand=(3,A) | | | | | | | | | | |
| 2HD7 | SC | GLN | 202 | A | Strand=(2,A) | SC | HIS | 237 | A | Strand=(3,A) | | | | | | | | | | |
| 2GIW | SC | ASN | 52 | A | Helix=(RH Alpha,2) | SC | THR | 78 | A | Coil | | | | | | | | | | |
| 2GD7 | SC | ASN | 53 | A | Helix=(RH Alpha,2) | SC | ASN | 53 | B | Helix=(RH Alpha,5) | | | | | | | | | | |
| 2FKI | SC | GLN | 101 | A | Helix=(RH Alpha,3) | SC | ASN | 105 | A | Helix=(RH Alpha,3) | | | | | | | | | | |
| 1Z1D | SC | ASN | 203 | A | Coil | SC | ASN | 210 | A | Helix=(RH Alpha,1) | | | | | | | | | | |
| 1W4U | SC | GLN | 34 | A | Strand=(1,AA) | SC | THR | 70 | A | Coil | SC | THR | 53 | A | Strand=(2,AA) | | | | | |
| 1QCE | SC | ASN | 37 | B | Helix=(RH Alpha,3) | SC | ASN | 82 | C | Helix=(RH Alpha,6) | | | | | | | | | | |
| | SC | ASN | 82 | A | Helix=(RH Alpha,2) | SC | ASN | 37 | C | Helix=(RH Alpha,5) | | | | | | | | | | |
| | SC | ASN | 37 | A | Helix=(RH Alpha,1) | SC | ASN | 82 | B | Helix=(RH Alpha,4) | | | | | | | | | | |
| 1OPZ | SC | THR | 57 | A | Helix=(RH Alpha,3) | SC | GLN | 61 | A | Helix=(RH Alpha,3) | | | | | | | | | | |
| 1M7L | SC | ASN | 77 | B | Coil | SC | ASN | 37 | A | Coil | SC | ASN | 117 | C | Coil | | | | | |
| 1ESX | SC | ASP | 7 | A | Coil | SC | ASP | 17 | A | Helix=(RH Alpha,1) | | | | | | | | | | |
| | BA | GLN | 3 | A | Coil | SC | GLN | 11 | A | Coil | SC | GLN | 8 | A | Coil | BA | ALA | 4 | A | Coil |
| 1EDL | SC | GLN | 24 | A | Helix=(RH Alpha,2) | SC | GLN | 53 | A | Coil | | | | | | | | | | |
| 1EDK | SC | GLN | 24 | A | Helix=(RH Alpha,2) | SC | GLN | 53 | A | Coil | | | | | | | | | | |
| 1EDI | SC | GLN | 24 | A | Helix=(RH Alpha,2) | SC | GLN | 53 | A | Coil | | | | | | | | | | |
| 1DPU | SC | ASN | 203 | A | Coil | SC | ASN | 210 | A | Helix=(RH Alpha,1) | | | | | | | | | | |
| 1A03 | BA | GLU | 86 | B | Coil | BA | GLY | 90 | B | Coil | BA | LYS | 89 | B | Coil | BA | ALA | 87 | B | Coil |

- Each row is a cycle. Up to 4 cooperative units were found per cycle.
- A PDB model may contain multiple cycles.
- BA/SC gives whether the unit is a backbone amide or a sidechain.
- The 3 character residue code is followed by the residue sequence number and chain ID.
- The type of a helix is identified, followed by its ID.
- For a beta strand, the strand number is given followed by the sheet ID.

The outstanding features of these data are how few cycles there are in the structures considered and, where present, how small these cycles are. When cycles consisting of just two sidechain amides are excluded, only 7 cycles remain, and only 3 of these are cycles of 4 potentially cooperative units. This striking result prompted review of the program used to generate the data, and the code used to determine HB connection between cooperative units was simplified to become the quite brief Haskell list comprehension mentioned above, but the results did not change. In the absence of formal proof of correctness or lesser corroboration, any program must be held to have the possibility of error. However, given the results of this program, it must be concluded that for standard proteins of size amenable to NMR spectroscopy i.e. less than ~30 kDa in mass, potentially cooperative HB cycles purely of protein are all but completely absent, so largely so that it is suggested that the absence of significant cycles is a fundamental property of protein structure. The potentially cooperative cycles that were detected serve as some test of the program's ability to detect at least cycles of amides. Note that the cycles detected by this program are not necessarily cooperative even though are comprised of potentially cooperative units. Determination of whether a HB cycle of potentially cooperative units is actually cooperative rests with quantum chemical analysis.

Two of the cycles that involve four potentially cooperative units, 1A03 and 1ESX, of Table 2 were isolated from their native protein context and were subject to quantum chemical energy minimization/geometry optimization at LC-wPBE(w=0.4)/6-311G** with TeraChem, retaining near-native geometry. Gaussian single point energy calculation of the resulting coordinates at the same method and basis were used for input to NBO 6.0 analysis. The 1A03 cycle is of backbone amides, and the 1ESX cycle is of backbone and sidechain amides. The HB and amide resonance data of these cycles appears in Table 3 and Table 4 and depiction of the geometries appears in Figure 21 and Figure 22. The extract of residues 83-90.B of PDB 1A03 has an HB between the backbone amides at the junctions of residues 84:85 and 87:88, but this is not involved in the cycle. This hydrogen bond is of length 1.829 angstroms with O(lp-s)->H-N* of 3.51 kcal/mol and O(lp-p)->H-N* less than 0.1 kcal/mol, and in total likely contributes to the stability of the structure of the extract. The tabled (Table 4) cycle of the extract of residues 2-12.A of PDB 1ESX contains a shorter cycle created by an HB with charge transfer from atom 10 (oxygen) to atom 95 (nitrogen) not shown in Table 4 but appearing in Figure 22, but its O(lp-s)->H-N* has associated energy of 0.47 kcal/mol and its O(lp-p)->H-N* 1.15 kcal/mol and so is weaker than the larger cycle.

The least HB in these cycles has very low energy charge transfers, and it may be concluded that the cyclic charge transfer is also very low. Only one backbone amide has a N(lp)->C-O(p)* value in keeping with RAHB for the method used, that being a 109.85 kcal/mol in the 1ESX extract. Thus, according to the quantum chemical methods used, these cycles are not cooperative. Consequently, the biological significance of these cycles is not investigated.

Table 3. Hydrogen Bonding Cycle in PDB Entry 1A03 extract of residues 83-90.B for LC-wPBE(w=0.4)/6-311G**

| Residue | N | N(lp)->C-O(s)* | N(lp)->C-O(p)* | O | O(lp-s)->H-N* | O(lp-p)->H-N* |
|---|---|---|---|---|---|---|
| BA(85,86) | 14 | 6.86 | 84.96 | 10 | 1.67 | 3.26 |
| BA(86,87) | 21 | 1.9 | 100.99 | 17 | 2.47 | 0.31 |
| BA(88,89) | 35 | 0.81 | 101.39 | 31 | 2.26 | 3.99 |
| BA(89,90) | 42 | 3.8 | 100.28 | 38 | 7.45 | 2.9 |

- BA(X,Y) : Backbone Amide formed at junction of residue numbered X and Y
- SC(X,Y) : Sidechain Amide of residue type X number Y
- N and O : Atom Ids of Amide Nitrogen and Oxygen in extract for electronic structure calculation
- Labels including "->" : Donor-Acceptor Interaction SOPT Energy in kcal/mol
- O Interaction is with H-N of N in the line below, O of last line to N of first line

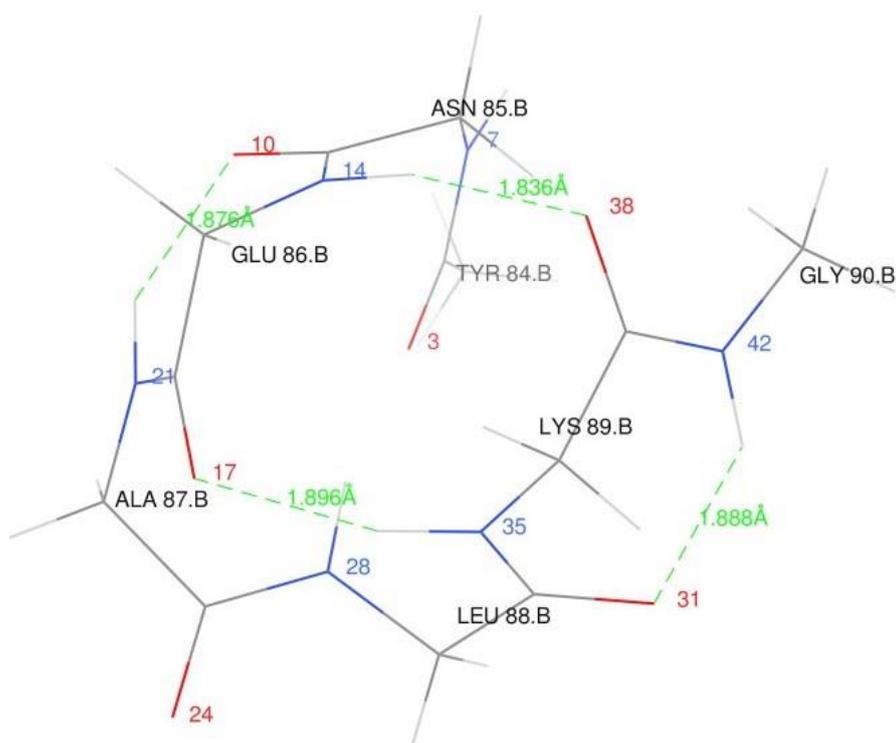

Figure 21. Hydrogen Bonding Cycle in PDB Entry 1A03

Table 4. Hydrogen Bonding Cycle in PDB Entry 1ESX extract of residues 2-12.A for LC-wPBE(w=0.4)/6-311G**

| Residue | N | N(lp)->C-O(s)* | N(lp)->C-O(p)* | O | O(lp-s)->H-N* | O(lp-p)->H-N* |
|---|---|---|---|---|---|---|
| BA(2,3) | 7 | 3 | 91.96 | 3 | 1.93 | 3.09 |
| BA(3,4) | 14 | 1.79 | 109.85 | 10 | 4.48 | 1.89 |
| SC(GLN,8) | 57 | 0.13 | 104.94 | 56 | 5.96 | 6.76 |
| SC(GLN,11) | 95 | 9.34 | 74.56 | 94 | 6.07 | 11.16 |

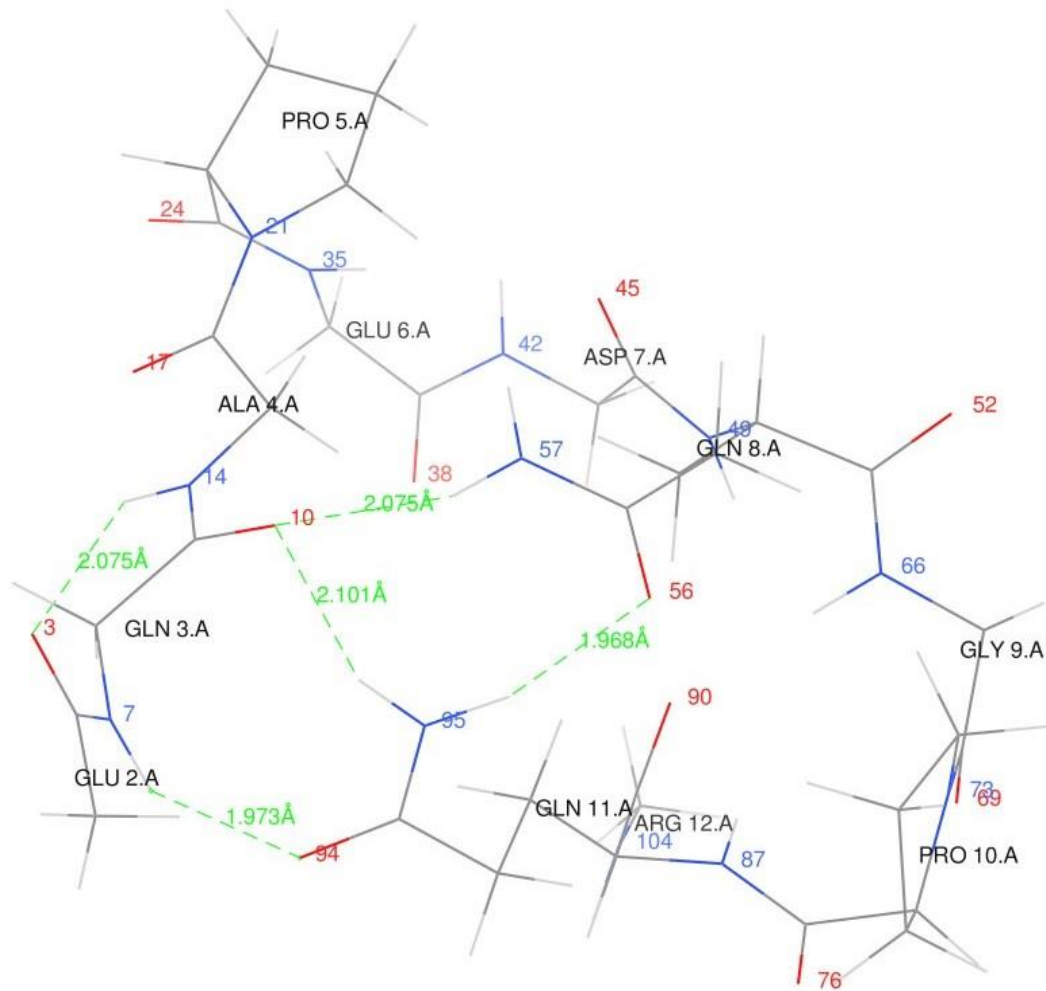

Figure 22. Hydrogen Bonding Cycle in PDB Entry 1ESX

## 5.8 Joined HB chains of beta sheet

Partial confirmation of the above finding was sought by querying the PDB extract without using detection of cycles. This extract was queried for asparagine or glutamine sidechain amides connecting any two beta sheet backbone amide HB chains without the requirement of cyclization, leaving detection of cycles to inspection. These queries do not require that asparagine and glutamine connect at the end of beta sheet HB chains, merely that they connect to beta sheet backbone amides. However, it is unlikely that a sidechain amide would bind to 2 beta sheet backbone amides other than at the ends of beta sheet HB chains, for such binding would weaken the beta sheet HB chains. These queries do not require that the beta sheet HB chains that are connected be in the same beta sheet. Such an asparagine or glutamine does not itself have to part of a beta sheet, or any other secondary structure. Each asparagine or glutamine sidechain amide was permitted to have multiple HB donating and accepting beta sheet backbone amides, but the results of the query were such that there were only one acceptor and donor under the defined criteria for HB which were the same as in the previous experiment.

These queries were applied to the same extract from the PDB as the previous experiment, and results were again striking. There were no instances of an asparagine sidechain amide connecting two beta sheet backbone HB chains, since there were no instances of an asparagine sidechain amide connecting two beta sheet backbone amides.

Only 18 instances of glutamine sidechain amides connecting beta sheet backbone amides exist in the models considered (Table 5). For a beta strand to be capped such that adjacent backbone amide RAHB chains of a beta sheet are HB connected by a sidechain amide, two configurations are possible, one in which the residue sequence numbers of the residues being HB connected do not differ, that is, the N-H and C-O are of the same residue, and the second configuration in which the residue sequence numbers of the beta strand differ by 2. 9 of these 18 instances are of the first configuration and 5 are of the second configuration. There are no instances of cycles being formed by sidechain amides and beta sheet backbone amides. The results of this experiment are consistent with the previous and more general experiment, giving some support to the view that no error exists in these programs.

Table 5. Glutamine sidechains connecting beta sheet backbone amides

| PDB | Res1 | Num1 | Ch1 | SS1 | Res2 | Num2 | Ch2 | SS2 | Res3 | Num3 | Ch3 | SS3 |
|---|---|---|---|---|---|---|---|---|---|---|---|---|
| 3AIT | GLN | 16 | A | Strand=(1,S1) | ASP | 58 | A | Strand=(3,S1) | ALA | 67 | A | Strand=(1,S2) |
| 2MA8 | GLN | 28 | A | Helix=(RH Alpha,1) | ILE | 75 | B | Strand=(2,B) | ALA | 77 | B | Strand=(2,B) |
| 2LQ7 | GLN | 421 | A | Strand=(3,A) | THR | 384 | A | Strand=(4,A) | GLU | 422 | A | Strand=(3,A) |
| 2LGR | GLN | 61 | A | Coil | VAL | 70 | A | Strand=(6,A) | VAL | 70 | A | Strand=(6,A) |
| 2LBT | GLN | 212 | A | Helix=(RH Alpha,1) | ILE | 323 | A | Strand=(3,A) | VAL | 325 | A | Strand=(3,A) |
| 2L7Q | GLN | 52 | A | Coil | GLY | 57 | A | Strand=(4,A) | GLY | 57 | A | Strand=(4,A) |
| 2KQM | GLN | 206 | B | Strand=(1,C) | TYR | 286 | B | Strand=(3,D) | CYS | 288 | B | Strand=(3,D) |
| 2KMX | GLN | 1177 | A | Strand=(4,A) | LYS | 1122 | A | Strand=(5,A) | TYR | 1178 | A | Strand=(4,A) |
| 2KMN | GLN | 131 | A | Helix=(RH Alpha,3) | ALA | 103 | A | Strand=(4,A) | ALA | 103 | A | Strand=(4,A) |
| 2KFS | GLN | 133 | A | Helix=(RH Alpha,7) | THR | 112 | A | Strand=(1,B) | THR | 112 | A | Strand=(1,B) |
| 2K3K | GLN | 70 | A | Helix=(RH Alpha,4) | ILE | 81 | A | Strand=(4,A) | ILE | 81 | A | Strand=(4,A) |
| 2CQO | GLN | 92 | A | Coil | TRP | 68 | A | Strand=(5,A) | TRP | 68 | A | Strand=(5,A) |
| 2BBI | GLN | 48 | A | Strand=(2,B) | CYS | 49 | A | Strand=(2,B) | ARG | 28 | A | Strand=(1,B) |
| 1WF9 | GLN | 40 | A | Helix=(RH Alpha,1) | SER | 22 | A | Strand=(1,A) | SER | 22 | A | Strand=(1,A) |
| 1QNZ | GLN | 118 | H | Strand=(1,D) | TYR | 206 | H | Strand=(2,E) | CYS | 208 | H | Strand=(2,E) |
|  | GLN | 6 | L | Strand=(1,A) | TYR | 90 | L | Strand=(1,C) | CYS | 92 | L | Strand=(1,C) |
| 1MKE | GLN | 94 | A | Strand=(1,A) | GLU | 95 | A | Strand=(1,A) | GLU | 95 | A | Strand=(1,A) |
| 1L2M | GLN | 31 | A | Helix=(RH Alpha,1) | VAL | 79 | A | Strand=(1,B) | VAL | 79 | A | Strand=(1,B) |

- The first outlined block identifies the residue with the connecting amide sidechain.
- The second outlined block identifies an HB accepting backbone amide.
- The third outlined block identifies an HB donating backbone amide.
- Multiple connections may occur with each PDB model.

Greater secondary structure stability is not necessarily biologically advantageous, but it seems unlikely that if greater stabilization were available that there would be no situation in which it would be advantageous, such as in proteins in which mechanical strength is crucial. Given this assumption and this PDB extract, it could be concluded that sidechain amide cyclization of beta sheet backbone HB is not stabilizing.

A hypothesis concerning stability of beta sheets was stated in [4], being that balance between the RAHB of the backbone amide HB chains favours the stability of beta sheets. Following this hypothesis, cooperative cyclization of a subset of the RAHB chains of a beta sheet would reduce stability.

In beta barrels such as [52] the RAHB chains are not cyclized, since the RAHB chains do not follow a diameter of the barrel. With sufficient extension of the length of the barrel by increasing the length of the beta strands, cyclization of some RAHB chains is then possible. Following the above hypothesis of the stability of beta sheets, this selective cooperative cyclization would reduce stability of the sheet. If examples of such long beta barrels exist in the PDB and were derived by X-ray crystallography [53], they would not have been detected by this survey which depends upon NMR to determine protonation state.

# 6   Conclusion

Unlike the lone pairs of the oxygen atom in water, the lone pairs of the carbonyl oxygen substantially maintain inequivalence in the presence of hydrogen bonding. One lone pair of carbonyl oxygen is p-type, has 2 lobes and is of higher energy than the s-rich lone pair. This is demonstrated by NBO analysis, and its calculations are not unitarily equivalent to a model in which the lone pairs are equivalent following sp2 hybridization [43, 44]. The higher energy of the p-type lone pair can donate more charge to an H-X* bond, depending on the geometric relation between the lobes of the p-type lone pair and the H-X* bond. This charge transfer is largest when the H-X* bond is in the plane of the oxygen p-type lone pair lobes and at a minimum when X is in the plane normal to that plane and passing through C and O.

This inequivalence of carbonyl oxygen lone pairs gives a better account of optimal C-O..N hydrogen bond angle, as an account with equivalent sp2 lone pairs could be expected to give optimal hydrogen bonds when the C-O..N angle is ~60 degrees from linear in the amide plane, whereas the shortest hydrogen bond length is found at ~75 degrees and is kept from optimality at angles yet further from linear by environmental steric interactions rather than intrinsic properties of the lone pairs. This geometry preference is not in keeping with a sp2 account of carbonyl oxygen lone pairs or a primarily dipolar account of HB.

Where classical molecular simulators model lone pairs of carbonyl/amide oxygen lone pairs, improvement in the accuracy of modelling could be had by capturing the inequivalence of the lone pairs, though this alone would not model amide resonance variation of the energy level and occupancy of these lone pairs.

A remarkable feature of amide-amide hydrogen bonding is that when the inter-amide steric exchange energy is deducted from the inter-amide donor-acceptor energy the result is close to zero, with slow increase as C-O..N becomes less linear in the plane of the amide having the C-O. The energy change associated with hydrogen bonding between amides primarily resides in variation of the resonance of the amides. The direct inter-amide NBO interactions are significant only in that they are a conduit for changes to the resonance of the amides. A particular consequence of this is that the majority of the energy associated with hydrogen bonding between amides is subject to electrostatic field via variation of amide resonance [4], increasing the significance of electrostatic field in determining protein structure.

The lobes of an amide oxygen p-type lone pair may concurrently engage in hydrogen bonding with different H-X* bonds. While this is anti-cooperative due to the busy donor effect, depending on C-O..N angles the total amount of charge donated from an amide is nonetheless substantially increased over the single hydrogen bond case. The increase in amide resonance due to both hydrogen bonds being present reduces the impact of the busy donor effect. Provided the two H-X* bonds are in the amide plane of the charge donating oxygen, the resonance stabilization can be greater than that for the case in which the oxygen participates in a single hydrogen bond, depending on the nature of resonant moieties extending from the hydrogen bonds.

While the focus of the search for useful materials has moved from HB structures such as polyamides to covalently bonded cylinders and sheets [51], it may be that useful HB materials may still be found with hydrogen bonded structures in which the lobes of the carbonyl p-type lone pair are used for separate or bifurcated hydrogen bonds, for the increase in total charge transfer means increased resonance and hence stability of the material, provided the carbonyl is part of a resonant system in the manner of the amide group. Constraints on the search for such materials were discussed above.

Cycles of coplanar amides become less energetically favourable per hydrogen bond as the number of amides in the cycle grows, for in the case of a regular C-O..N angle this angle becomes closer to linear, and in the case of irregular C-O..N angles the cooperativity is limited by the most linear of these angles. The remaining possibility is for the amides to not be co-planar, which does not mean that hydrogen bonding is not suboptimal, for all that is required is for the H-N to be in the plane of the amide bearing the charge donating oxygen, which allows the amide bearing the H-N to be rotated about its H-N axis since the H-N orbitals are symmetric about that axis. This has the additional possible benefit of reducing steric conflict introduced by amide substituents such as the CA atoms and its substituents. We have not proposed a cyclic example of this configuration, nor suggested a regular bending of backbone chains that would accomplish this. Bearing in mind the limitation of the cooperativity to the least cooperation in the cycle, a regular bend of the backbone seems necessary for a cycle of backbone amides. In a mixed cycle of backbone and sidechain amides, the lower resonance of the sidechain amides will be limiting. Moieties with less hydrogen bonding cooperativity than amides introduce further limitation of cyclic cooperativity.

Cooperativity of cyclic HB can be diminished through the busy donor effect where an acceptor is extraneous to the cycle. A feature of RAHB secondary structures is that the backbone amide oxygens are largely protected from extraneous interactions. This protection is an additional burden on any regular backbone bend such as mentioned in the previous paragraph. Arrangements of non-planar cycles with regular backbone bends such that cycles can be mutually supportive and protecting are yet more constraining.

Our tentative finding is that cooperative HB cycles in standard proteins are radically disfavoured in nature. The use of data deriving from NMR spectroscopy biases the study toward smaller proteins, but as small cycles can be expected to be more frequently occurring than large cycles, no potentially cooperative cycles of more than 4 units being detected in the selected NMR data and the 2 4-unit examples investigated by quantum chemical means revealed to not be cooperative at the level of theory used suggest that cooperative cycles of hydrogen bonding of more than 3 units do not exist PDB structures. This observation rests on detection by program of potentially cooperative hydrogen bonded cycles in the PDB. Since the programs used for this detection have not been formally verified as solving the problem to be solved, independent confirmation of this finding is highly desirable.

Possible causes of the near-complete absence of these cycles additional to the foregoing arise from dynamics. In a cooperative HB cycle there is one HB chain, unlike the multiple spines of RAHB helices or RAHB chains of beta sheets which serve to stabilize geometry including HB lengths of the secondary structure. The single chain of a cycle can be expected to be susceptible to thermal geometry variation. Whereas RAHB secondary structures tend to be islands of stability in protein geometry fluctuation, the HB of cooperative cycles will transiently suffer stretching, and cooperativity is lost rapidly with increased HB length due to rapid reduction in charge transfer with distance. This local loss of cooperativity will propagate around the cycle, attenuating the cooperativity benefit of the cycle.

A hypothesis of the stability of beta sheets [4] proposes that selective enhancement of RAHB in some backbone amide HB chains of a beta sheet reduces the stability of the sheet, which disfavours cooperative cyclization of some HB chains of a beta sheet. We are presently unable to investigate this proposal due to pronounced errors of established DFT methods as applied to beta sheets [11].

Cyclic cooperativity of HB is fragile even in a non-dynamic calculation. The cooperativity of the cycle is limited to the least cooperative unit of the cycle, and the cooperativity at each unit is geometry sensitive. In dynamics, the entropic penalty for maintenance of geometry conducive to cooperativity for all units in the cycle will be high. Acyclic network cooperativity does not have the particular cooperativity advantage of cyclic cooperativity or its fragility. While inter-peptide HB secondary structures such as alpha helices and beta sheets are extensible without loss of the cooperativity already established, extension of a cycle involves breaking the cycle, loss of all cyclic cooperativity and change to the geometry of the cycle to allow a new element to be inserted into the cycle.

It may be that cooperative cycles result in lower density of stabilizing cooperativity than is possible in non-cyclic cooperative structures.

It may also be that evolution has selected in protein a structurally specific polymer to disfavour cooperative cycles so that energetic equality between HB cycles and non-cyclic HB chains does not lead to ambiguity in what fold a given residue sequence specifies. Where modification of the standard protein backbone, such as for beta-peptides [54], or introduction of artificial amino-acid residues results in non-specific structure, it might be investigated whether cooperative cycles of HBs are present.

# 7 Acknowledgements

Prof. John A. Carver is acknowledged for reading this manuscript and offering editing suggestions.

eResearch South Australia is acknowledged for hosting and administering machines provided under Australian Government Linkage, Infrastructure, Equipment and Facilities grants for Supercomputing in South Australia, directing funds to the acquisition of Nvidia Tesla GPU nodes and allocating 64 CPU cores and 256 GB RAM of the NeCTAR Research Cloud (a collaborative Australian research platform supported by the National Collaborative Research Infrastructure Strategy) to the present work.

# 9 Appendix 1

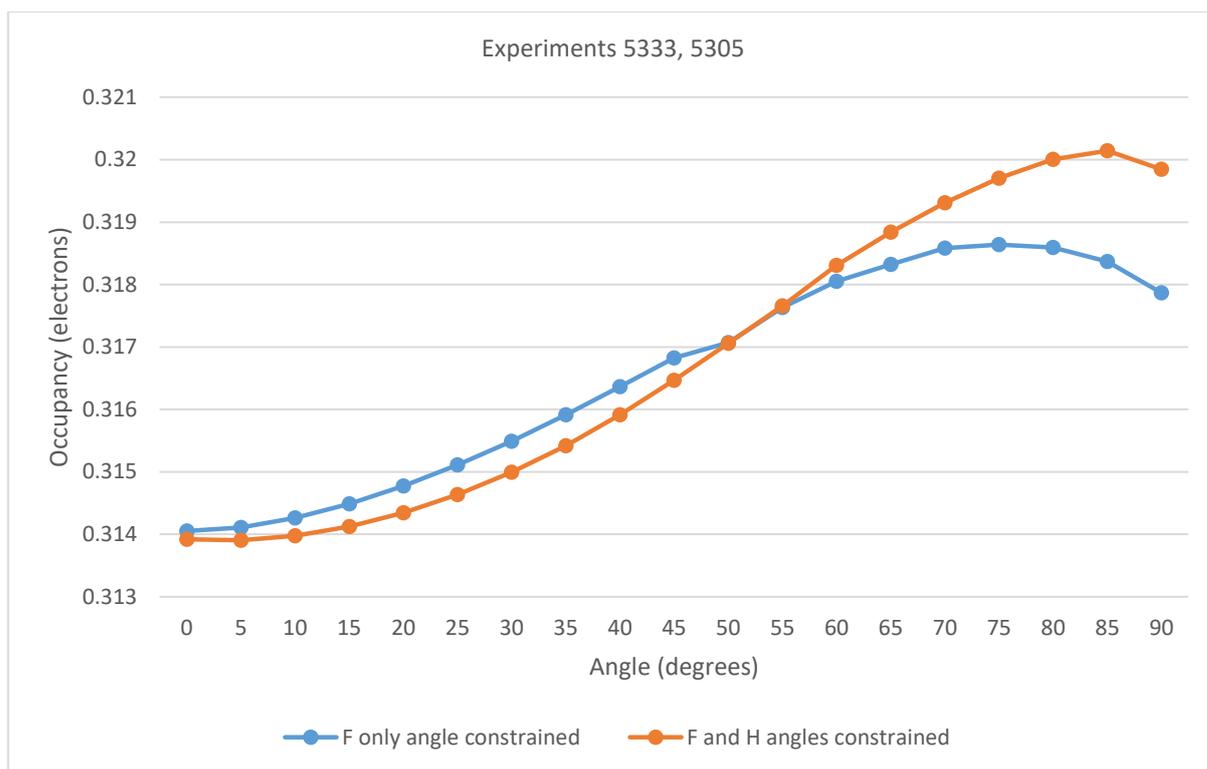

Figure 23. C-O(p)* NBO Occupancy in N-methylformamide with HF Hydrogen Bonded and F only or H and F Constrained to Angle from C-O at O in Amide Plane with SCS-MP2/aug-cc-pVTZ

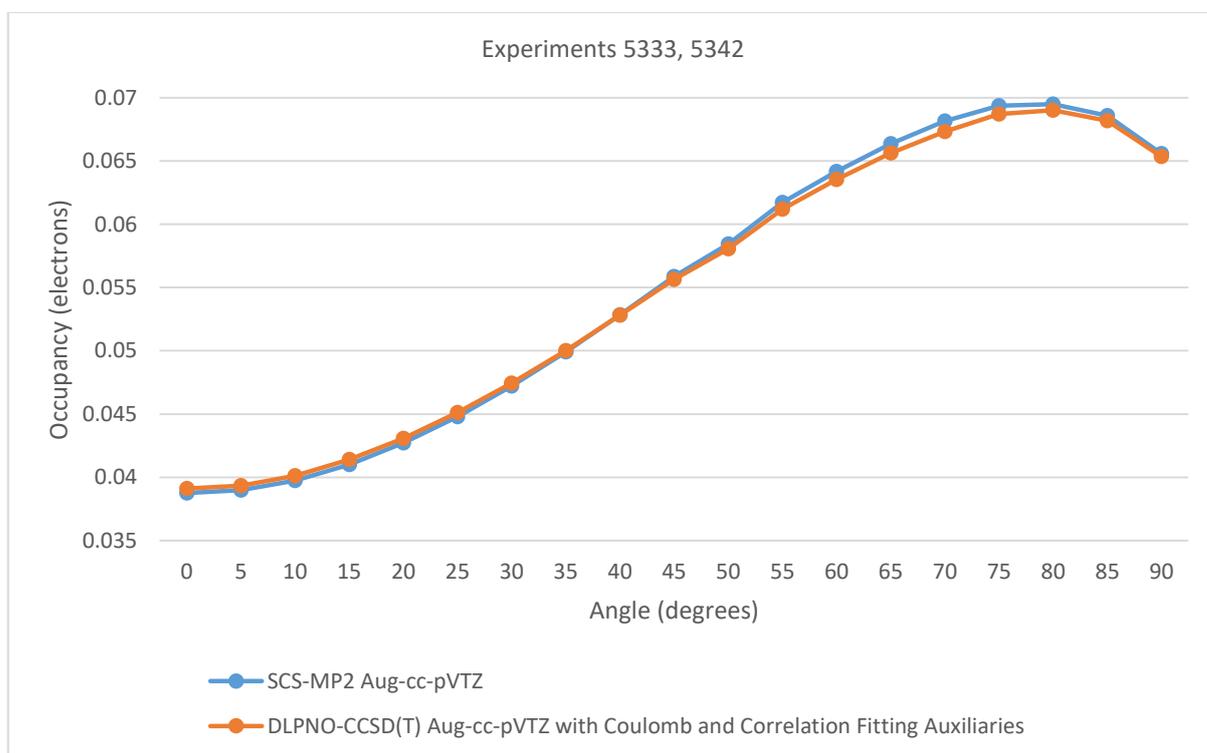

Figure 24. H-F* NBO Occupancy with HF Hydrogen Bonded to N-methylformamide O at C-O-F Angle in Amide Plane with Geometry Optimized at SCS-MP2/aug-cc-pVTZ

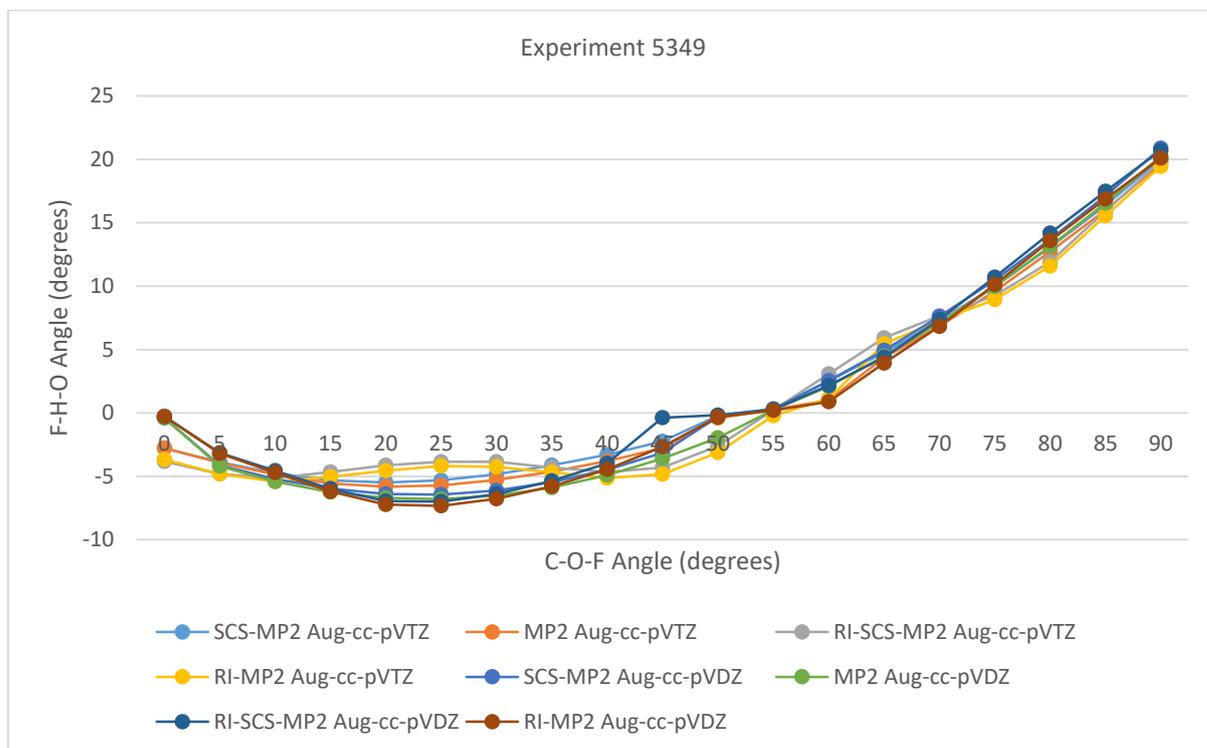

Figure 25. F-H-O Angle at C-O-F Angle in HF Hydrogen Bonded to N-methylformamide O in Amide Plane, F-H Pointing Outboard of O as Positive Angle and Between C and O as Negative Angle. RI Methods Used with Coulomb and Correlation Auxiliary Basis Sets

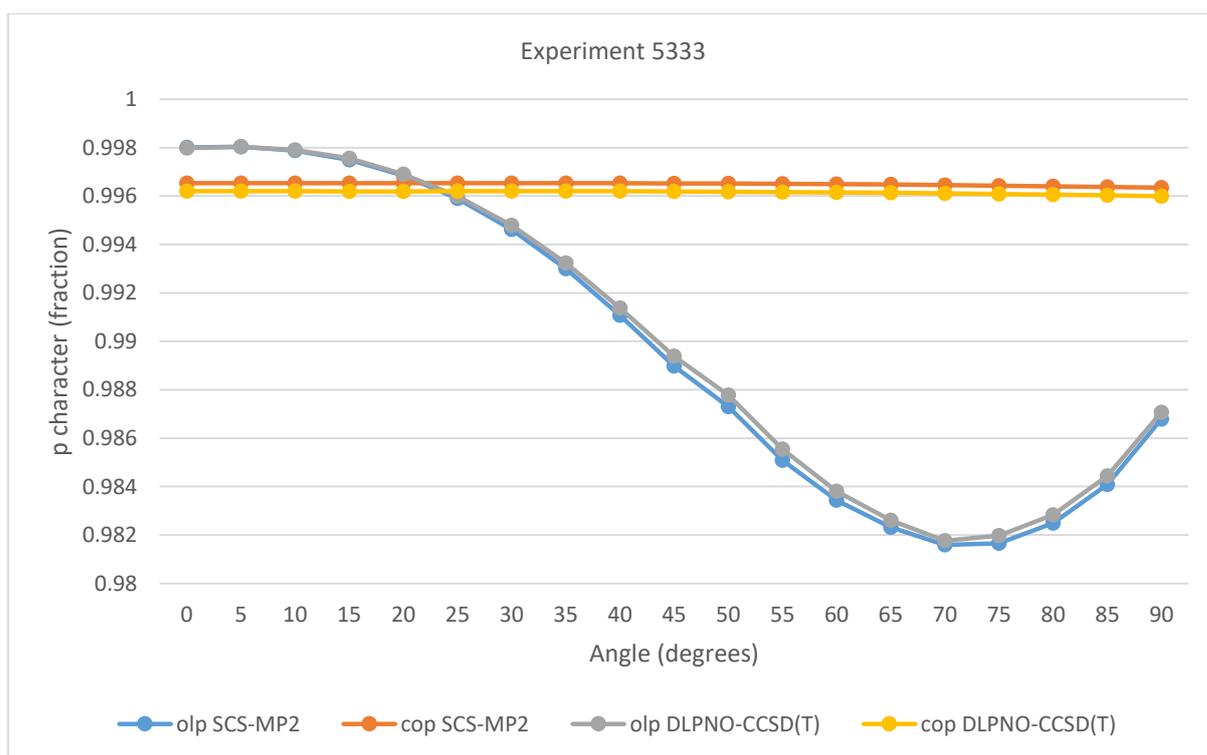

Figure 26. O(lp-p) and C-O(p) NBO p Character with N-methylformamide O Hydrogen Bonded to HF at C-O-F Angle in Amide Plane with F Distal to N at SCS-MP2/aug-cc-pVTZ Optimized Geometry and DLPNO-CCSD(T)/aug-cc-pVTZ with Coulomb and Correlation Auxiliaries

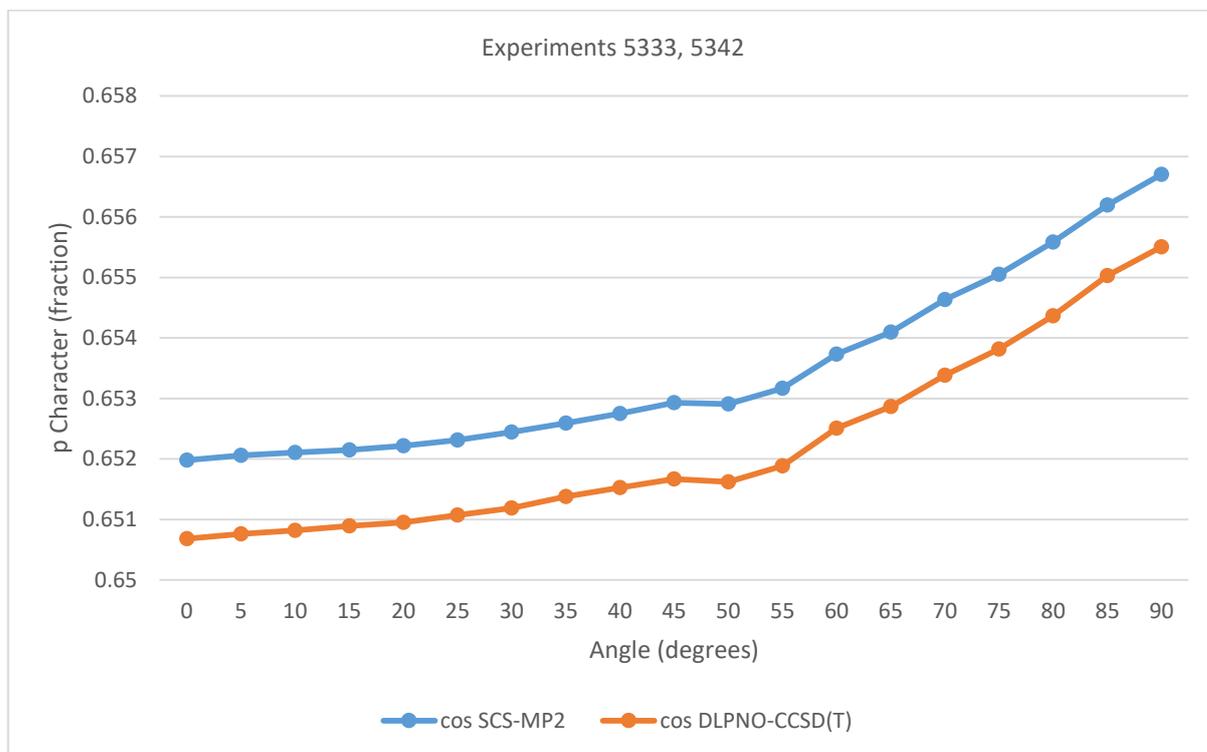

Figure 27. C-O(p) NBO p Character in N-methylformamide Hydrogen Bonded to H-F at C-O-F Angle in Amide Plane with F Distal to N at SCS-MP2/aug-cc-pVTZ Optimized Geometry and DLPNO-CCSD(T) with aug-cc-pVTZ and Coulomb and Correlation Auxiliaries

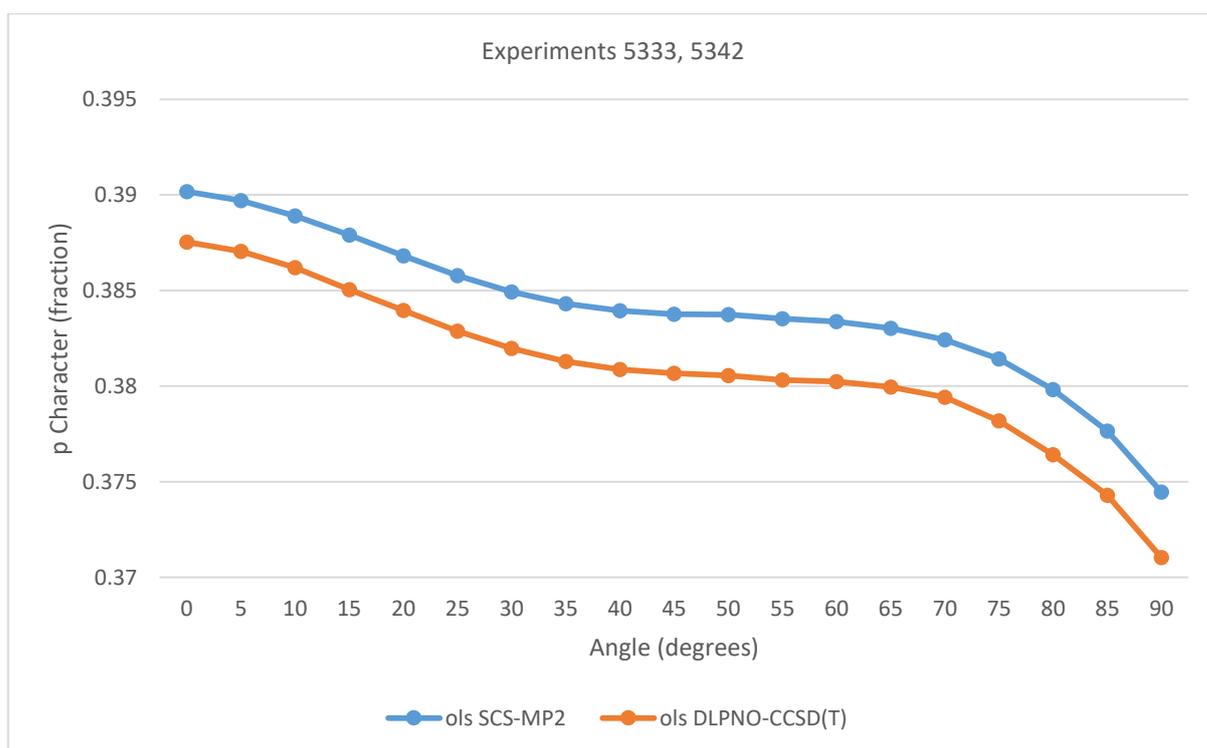

Figure 28. O(lp-s) NBO p Character in N-methylformamide Hydrogen Bonded to H-F at C-O-F Angle in Amide Plane with F Distal to N at SCS-MP2/aug-cc-pVTZ Optimized Geometry and DLPNO-CCSD(T)/aug-cc-pVTZ with Coulomb and Correlation Auxiliary Basis Sets

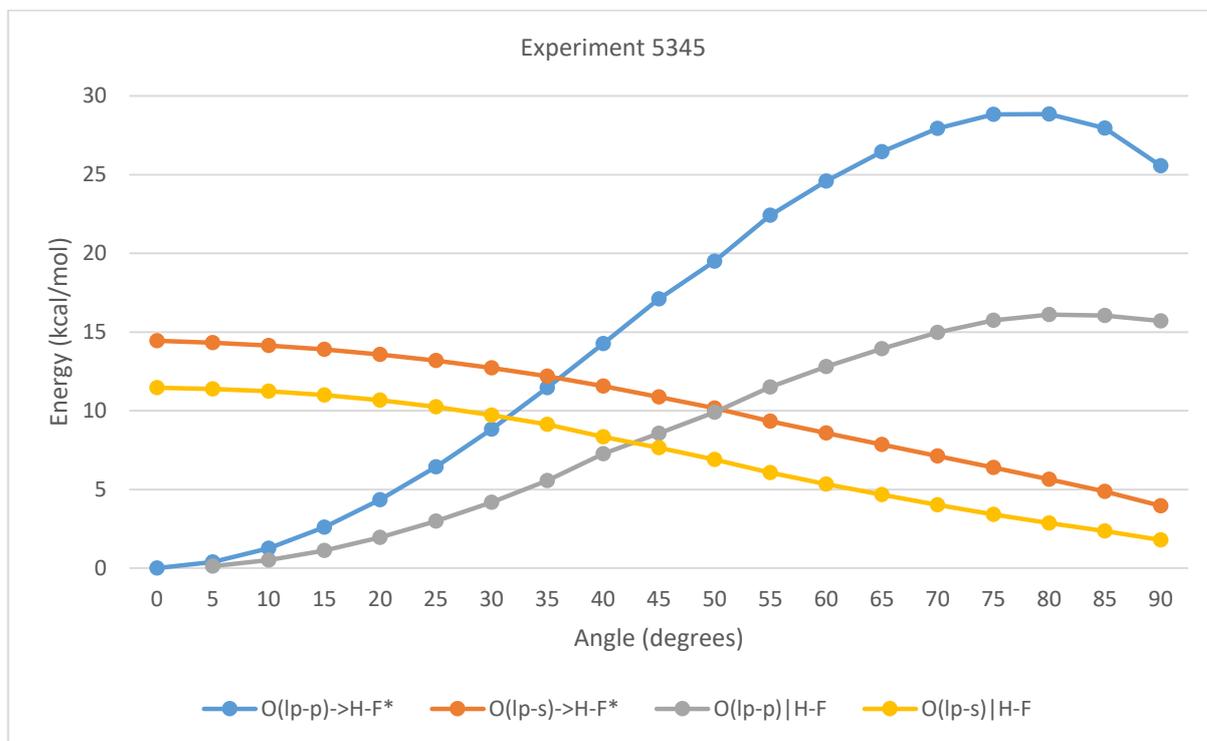

Figure 29. Donor-Acceptor SOPT and Steric Exchange Energies in N-methylformamide O Lone Pair Interactions with H-F and H-F* at C-O-F Angle with Geometry Optimized at SCS-MP2/aug-cc-pVTZ with Energetics at LC-wPBE(w=0.4)/aug-cc-pVTZ

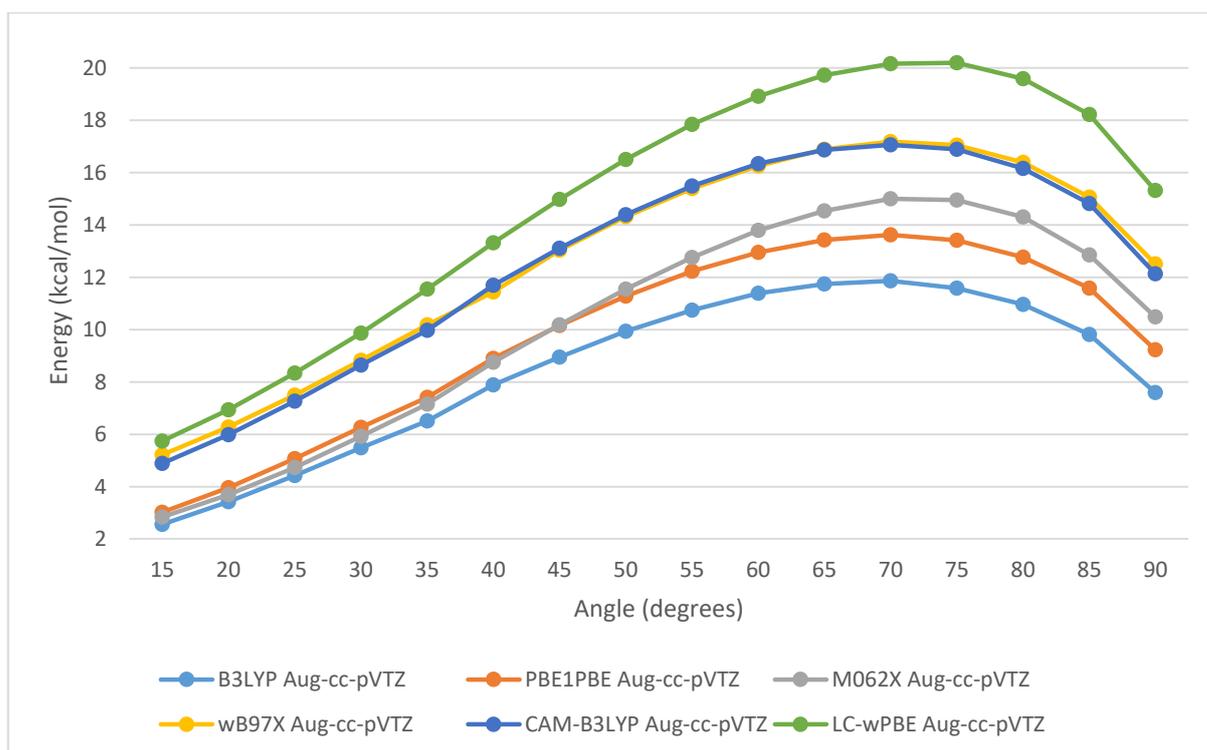

Figure 30. Donor-Acceptor SOPT Energy Minus Steric Exchange Energy for Interactions Between N-methylformamide O Lone Pairs with H-F and H-F* NBOs at C-O-F Angle in Amide Plane with Geometry Optimized at Same Method

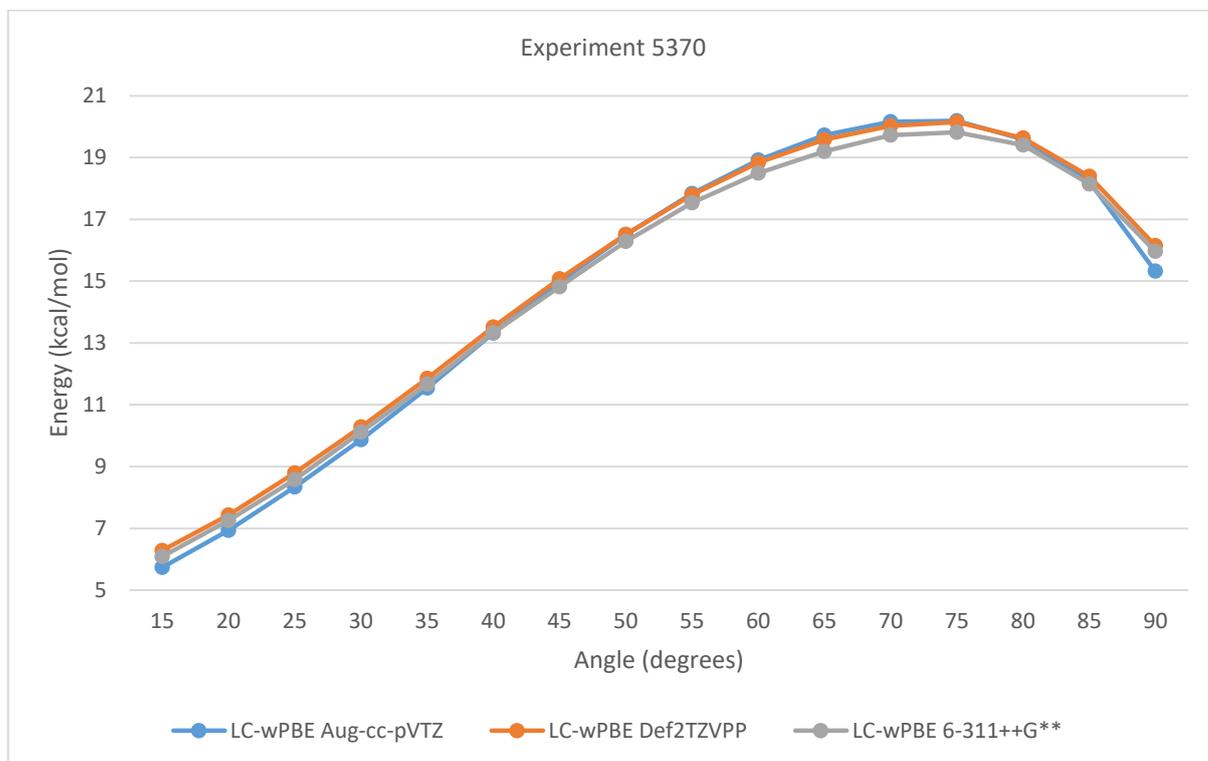

Figure 31. Donor-Acceptor SOPT Minus Steric Exchange Energy for N-methylformamide O Lone Pairs Interaction with H-F and H-F* NBOs at Varying C-O-F Angle in Amide Plane at 3 Basis Sets

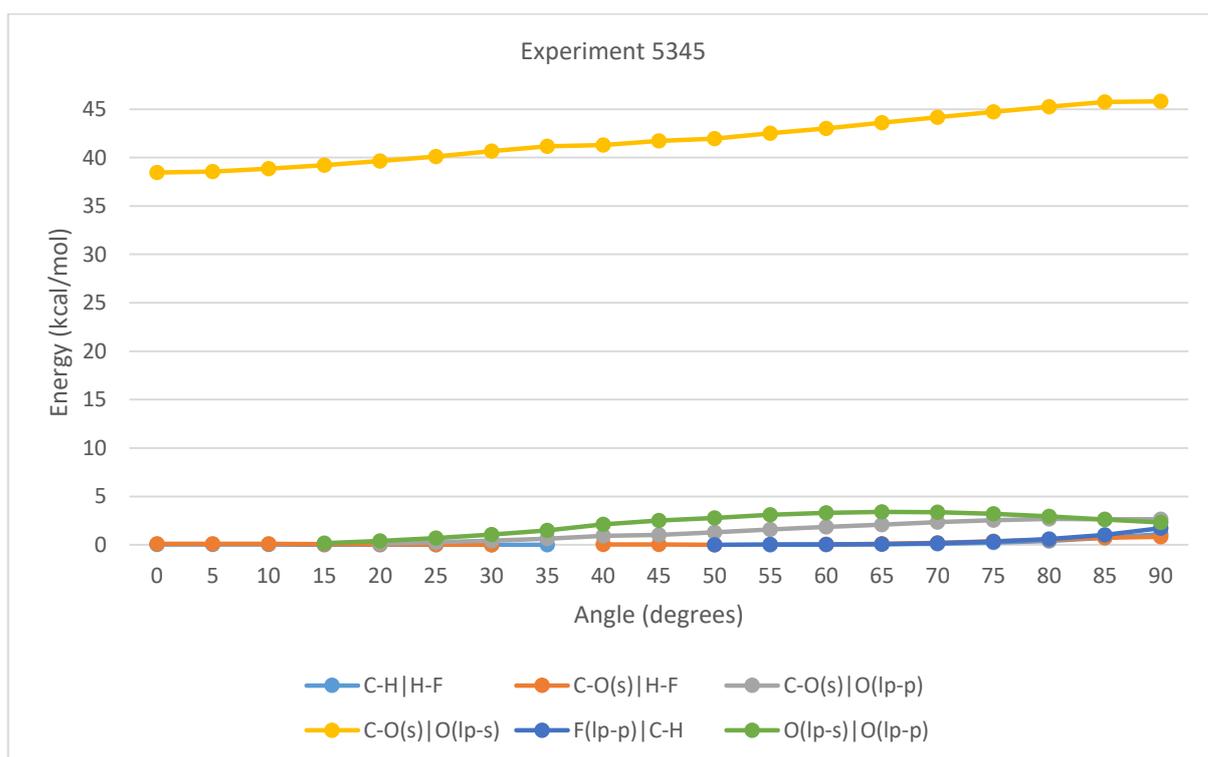

Figure 32. Steric Exchange Energy of Interactions Not Directly Involved in Hydrogen Bonding of HF to N-methylformamide O at C-O-F Angle in Amide Plane with Geometry Optimized as SCS-MP2/aug-cc-pVTZ and Energetics at LC-wPBE(w=0.4)/aug-cc-pVTZ

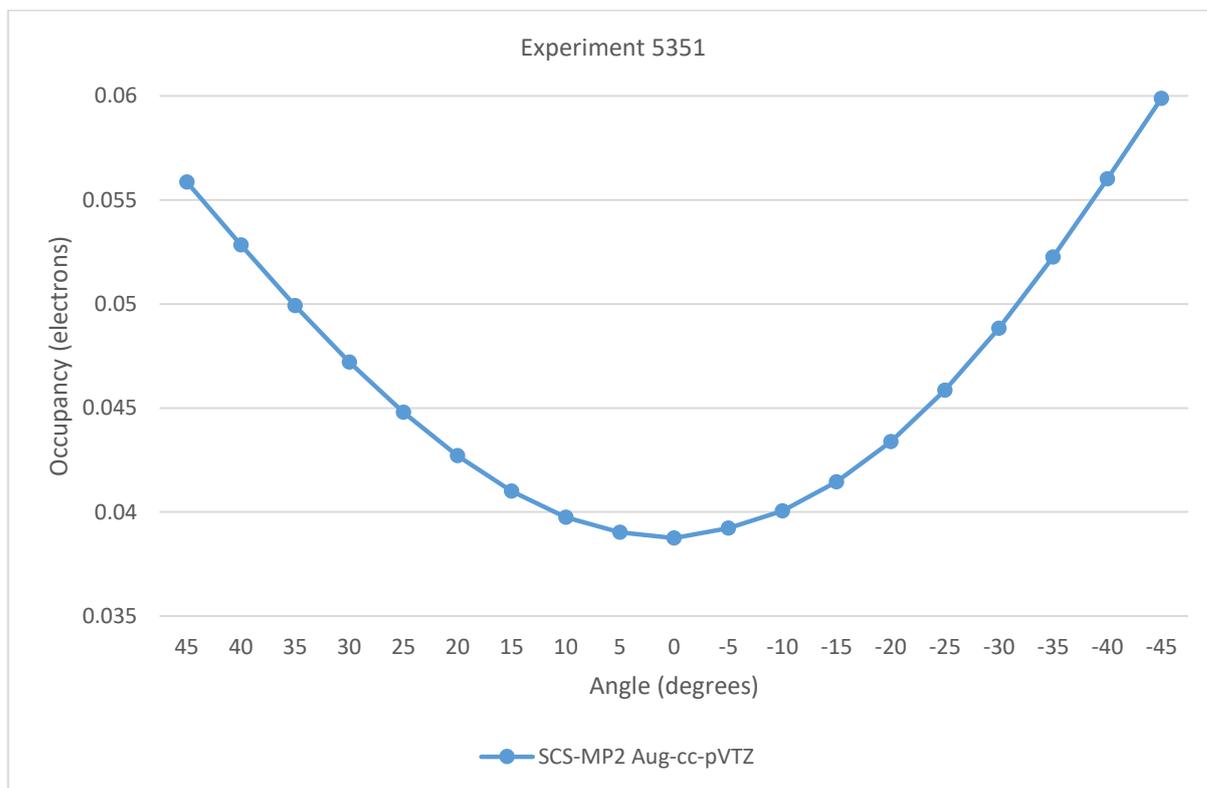

Figure 33. H-F* NBO Occupancy in HF Hydrogen Bonded to N-methylformamide O at C-O-F Angle in Amide Plane with Angles Distal to N being Positive

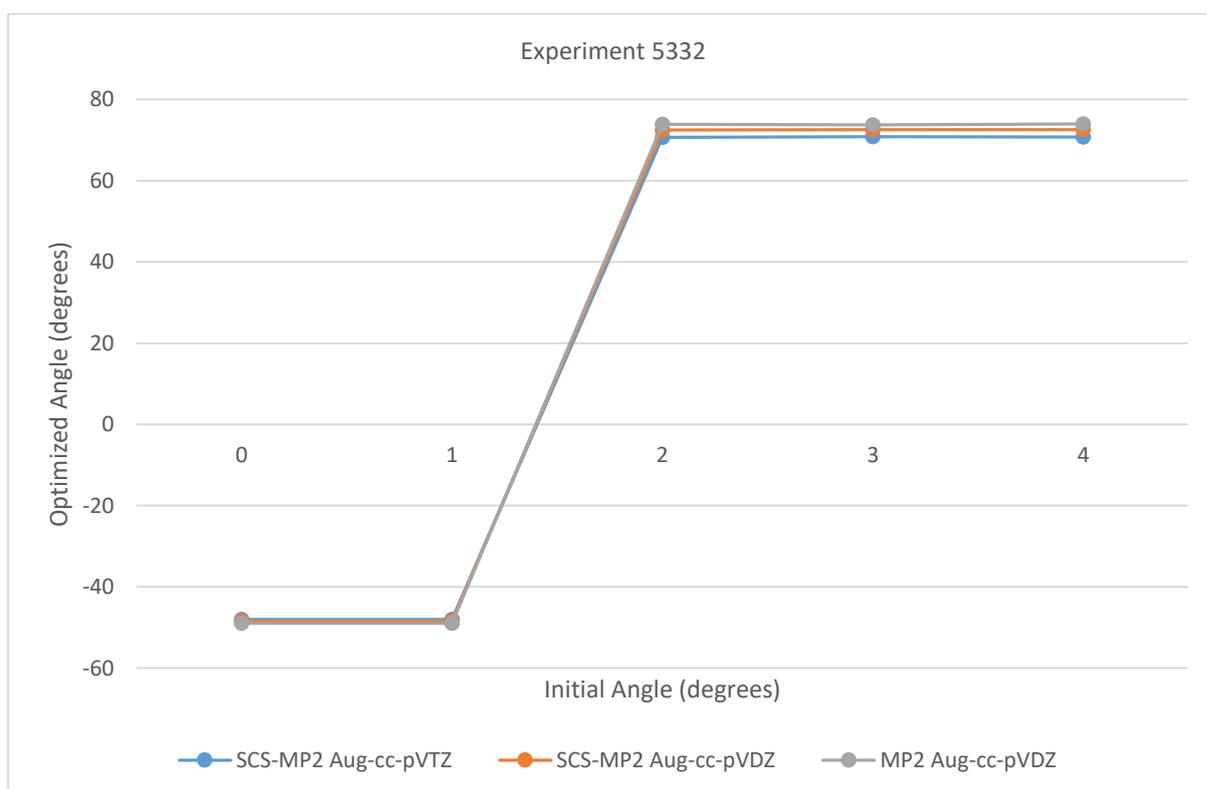

Figure 34. Initial versus Geometry Optimized C-O-F Angle in Amide Plane in HF Hydrogen Bonded to N-methylformamide O, Negative Angles for F Proximal to N

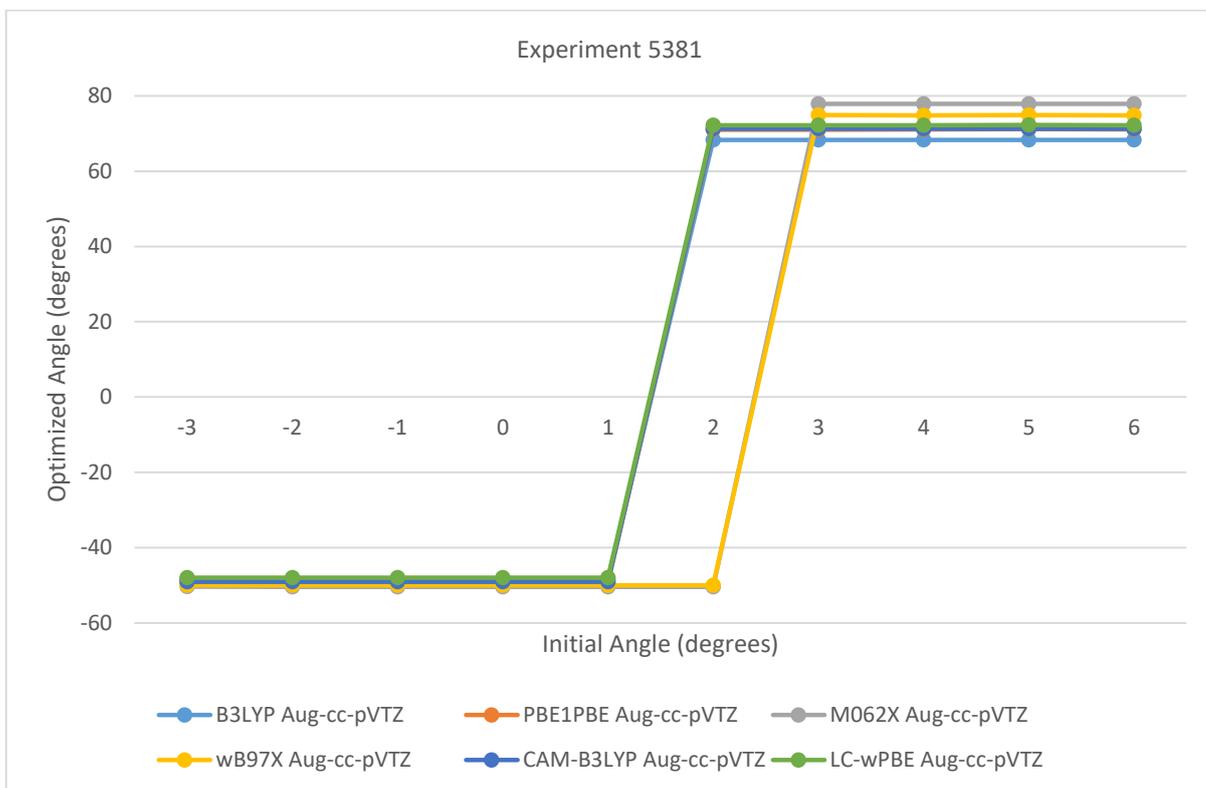

Figure 35. Initial versus Geometry Optimized Angle in HF Hydrogen Bonded to N-methylformamide O, Negative Angles for F Proximal to N

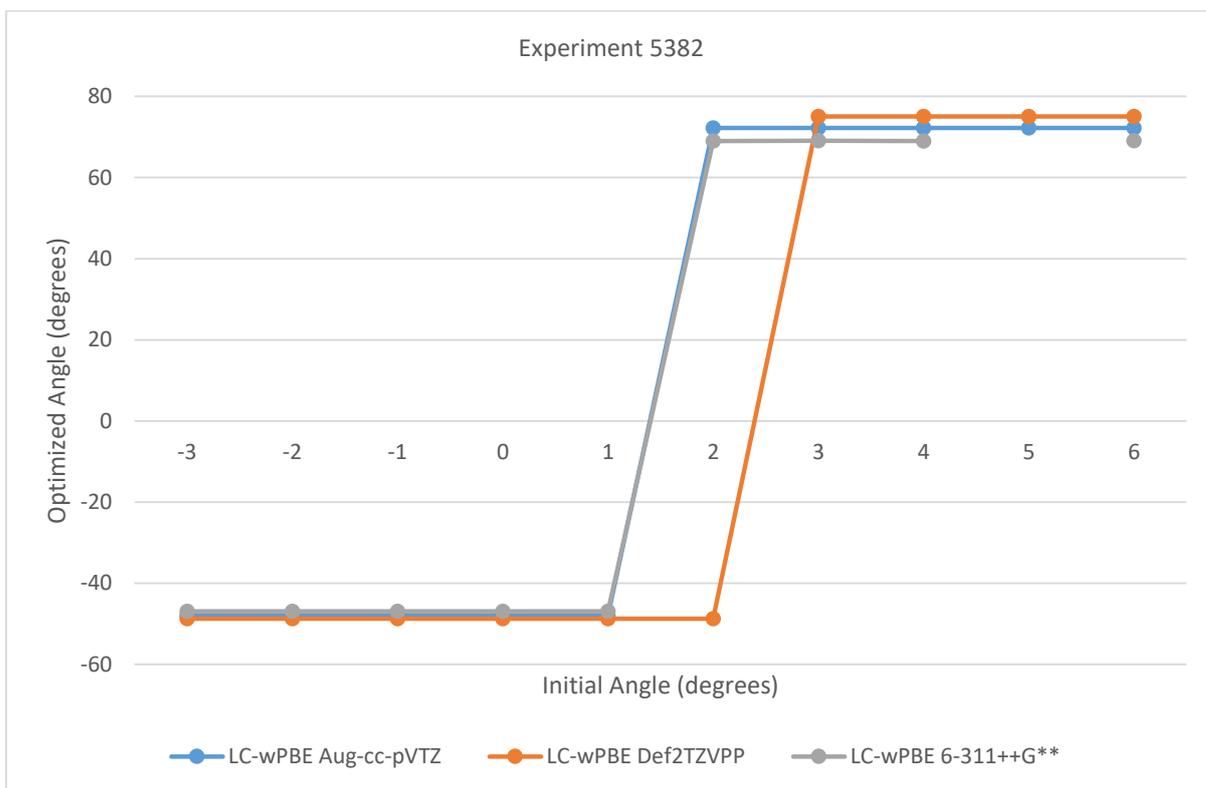

Figure 36. Initial versus Geometry Optimized C-O-F Angle with HF Hydrogen Bonded to N-methylformamide O, Negative Angles for F Proximal to N

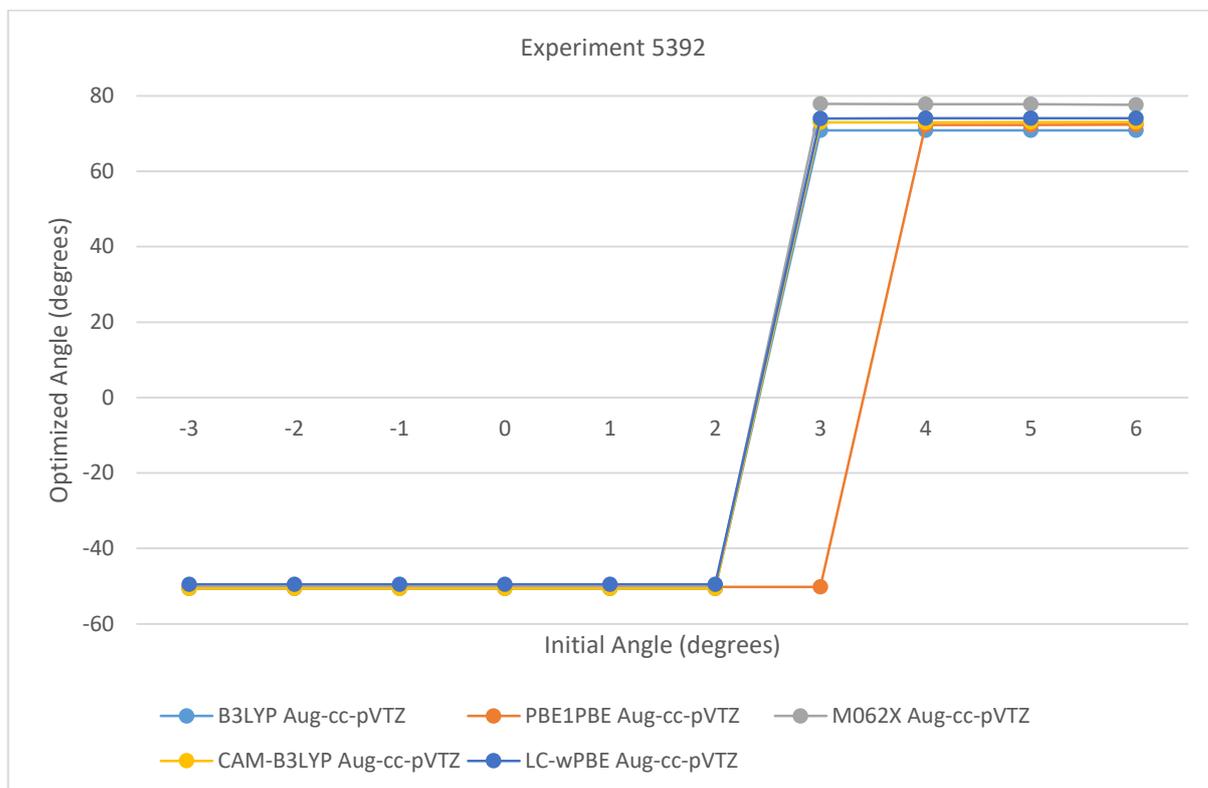

Figure 37. Initial versus Optimized C-O-F Angle with HF Hydrogen Bonded to N-methylformamide O in Amide Plane, Negative Angle for F Proximal to N, All Methods with D3 Dispersion Correction

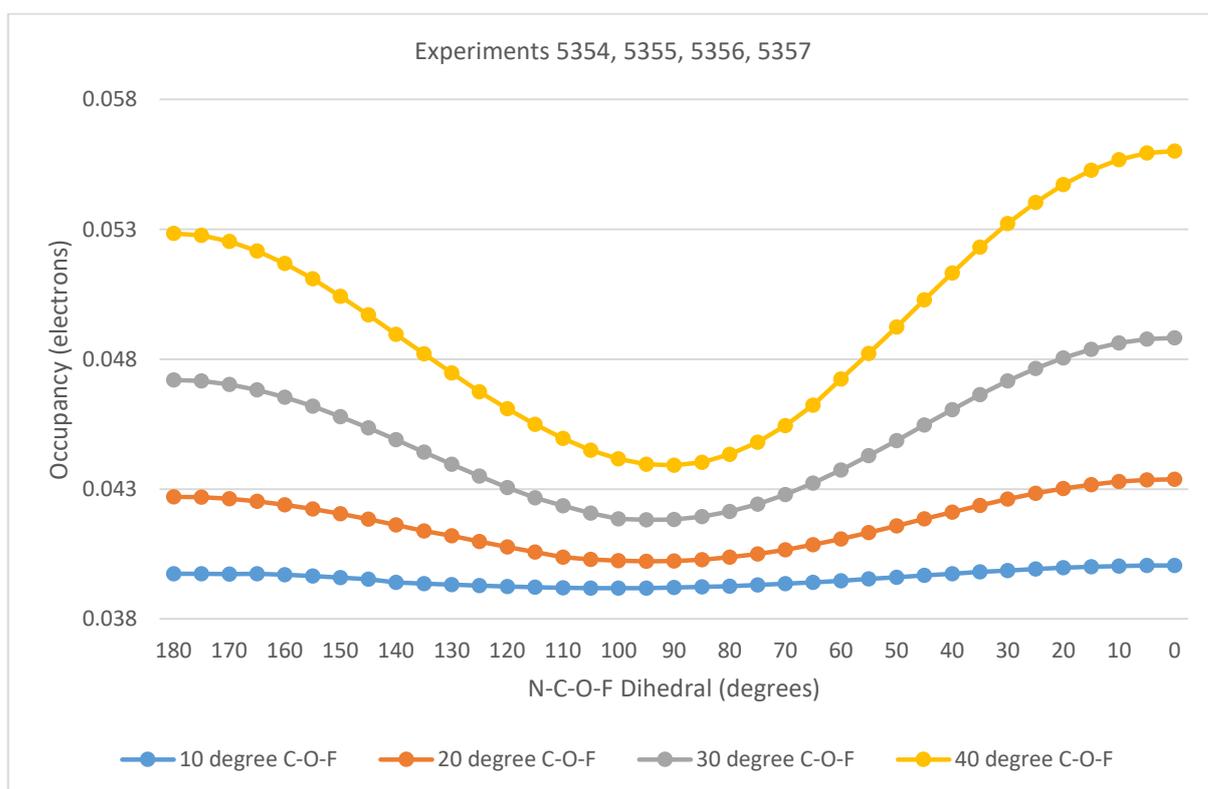

Figure 38. H-F* NBO Occupancy HF Hydrogen Bonded to N-methylformamide O at Constant C-O-F Angle with F Rotated About C-O Axis at SCS-MP2/aug-cc-pVTZ

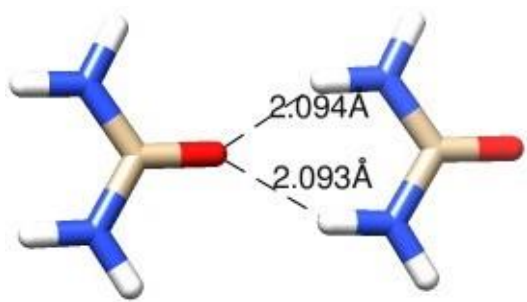

Figure 39. Urea Hydrogen Bonded with Urea at SCS-MP2/aug-cc-pVTZ

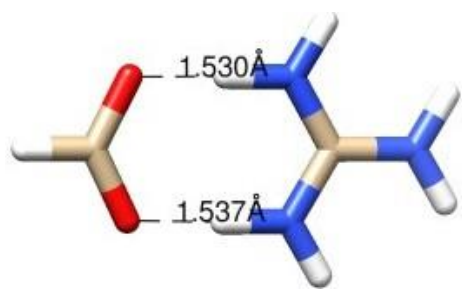

Figure 40. Formate Hydrogen Bonded with Guanidinium at SCS-MP2/aug-cc-pVTZ

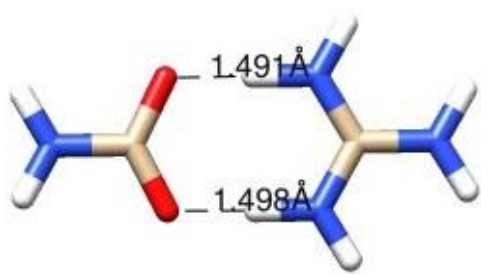

Figure 41. Carbamate Hydrogen Bonded with Guanidinium at SCS-MP2/aug-cc-pVTZ

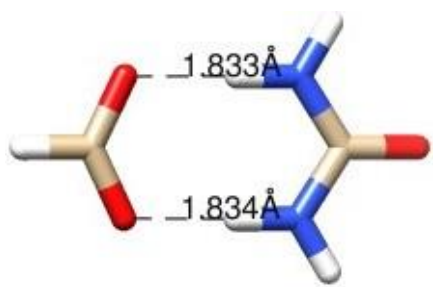

Figure 42. Formate Hydrogen Bonded with Urea at SCS-MP2/aug-cc-pVTZ

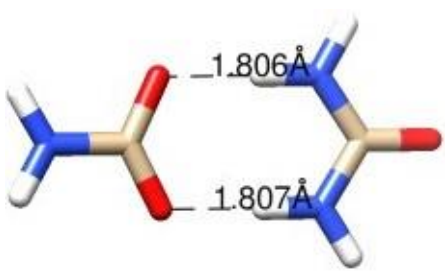

Figure 43. Carbamate Hydrogen Bonded with Urea at SCS-MP2/aug-cc-pVTZ

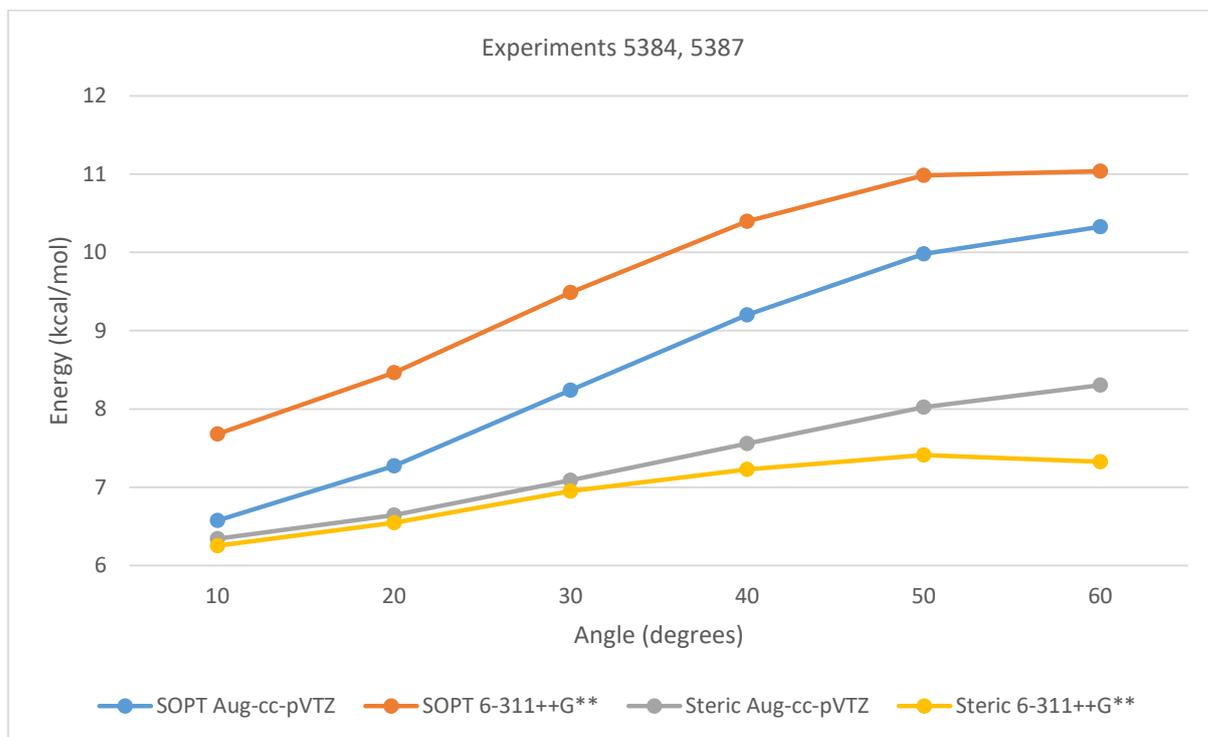

Figure 44. Donor-Acceptor SOPT and Steric Exchange Energies for Interactions Between Hydrogen Bonded N-methylformamides at Given C1-O1-N2 Angle with N1-C1-O1-N2 Dihedral 180 Degrees at LC-wPBE(w=0.4)

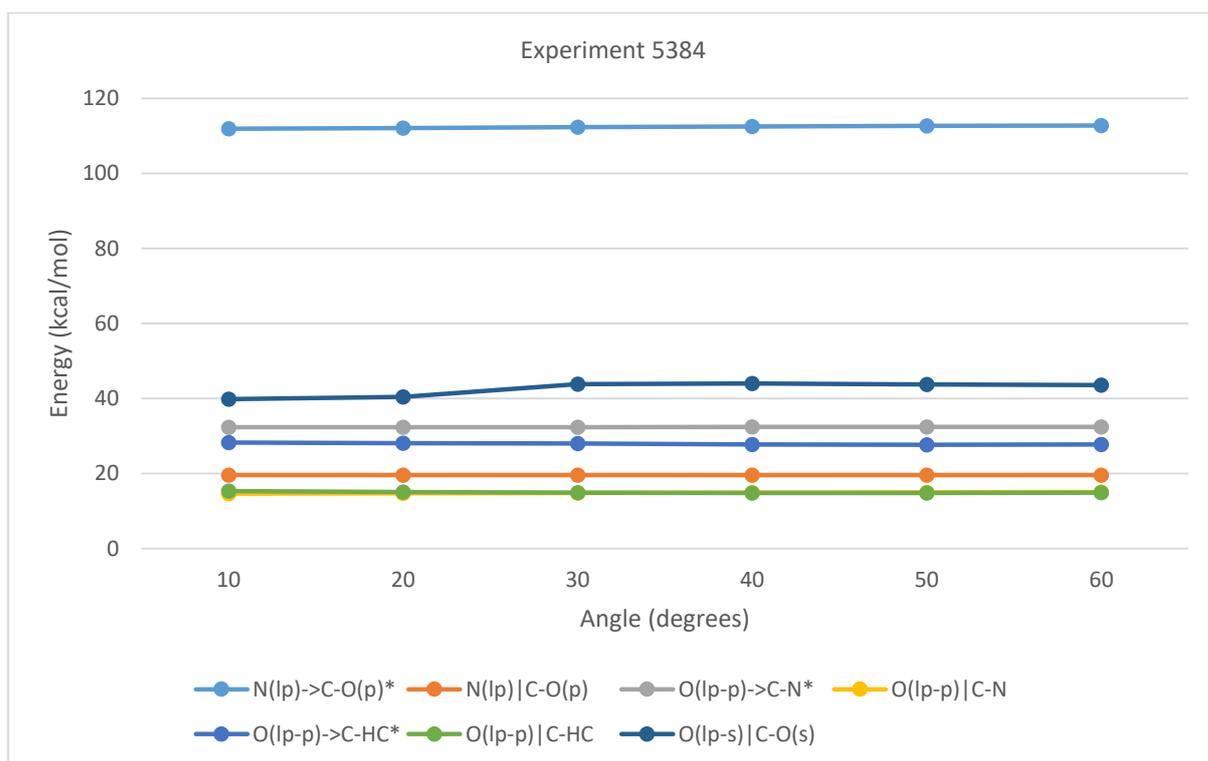

Figure 45. Selected Donor-Acceptor SOPT and Steric Exchange Energies for Amide 1 having O Involved in Hydrogen Bonding Between Pair of N-methylformamides at Given C-O-N Angle in Common Amide Plane at LC-wPBE(w=0.4)/aug-cc-pVTZ

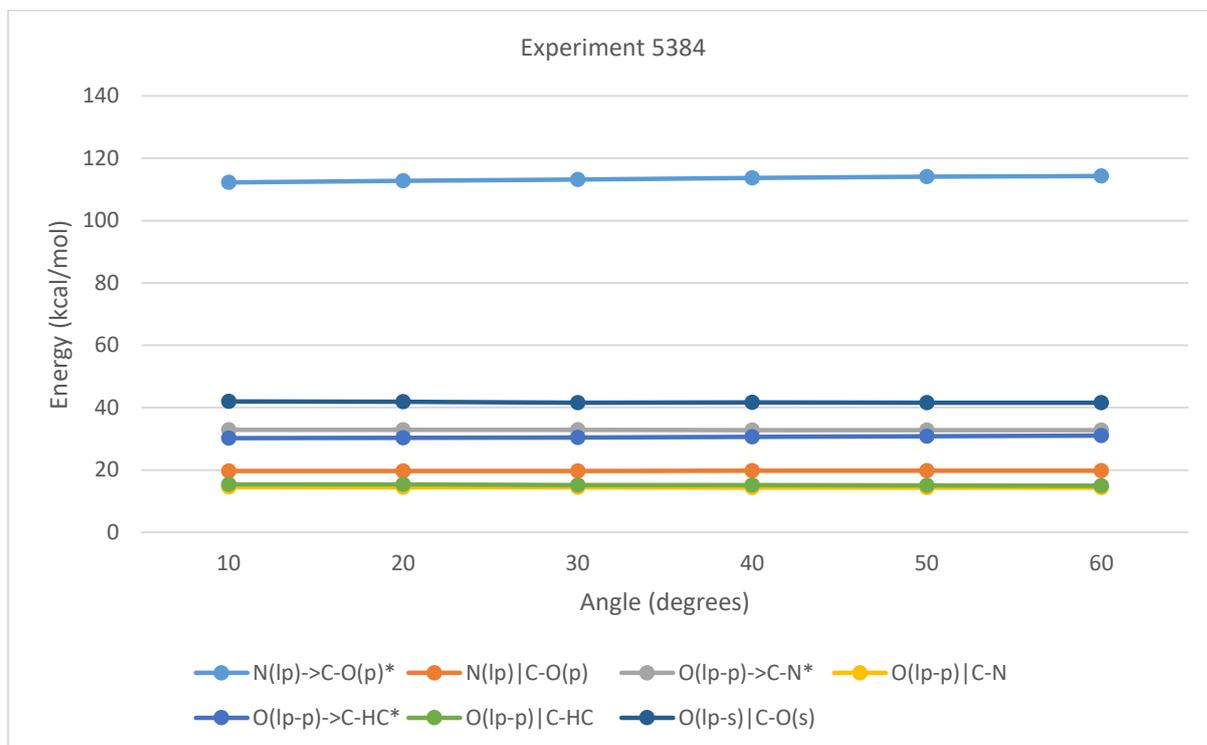

Figure 46. Selected Donor-Acceptor SOPT and Steric Exchange Energies in Amide 2 having H-N involved in Hydrogen Bonded Pair of N-methylformamides at Given C-O-N Angle in Common Amide Plane at LC-wPBE(w=0.4)/aug-cc-pVTZ

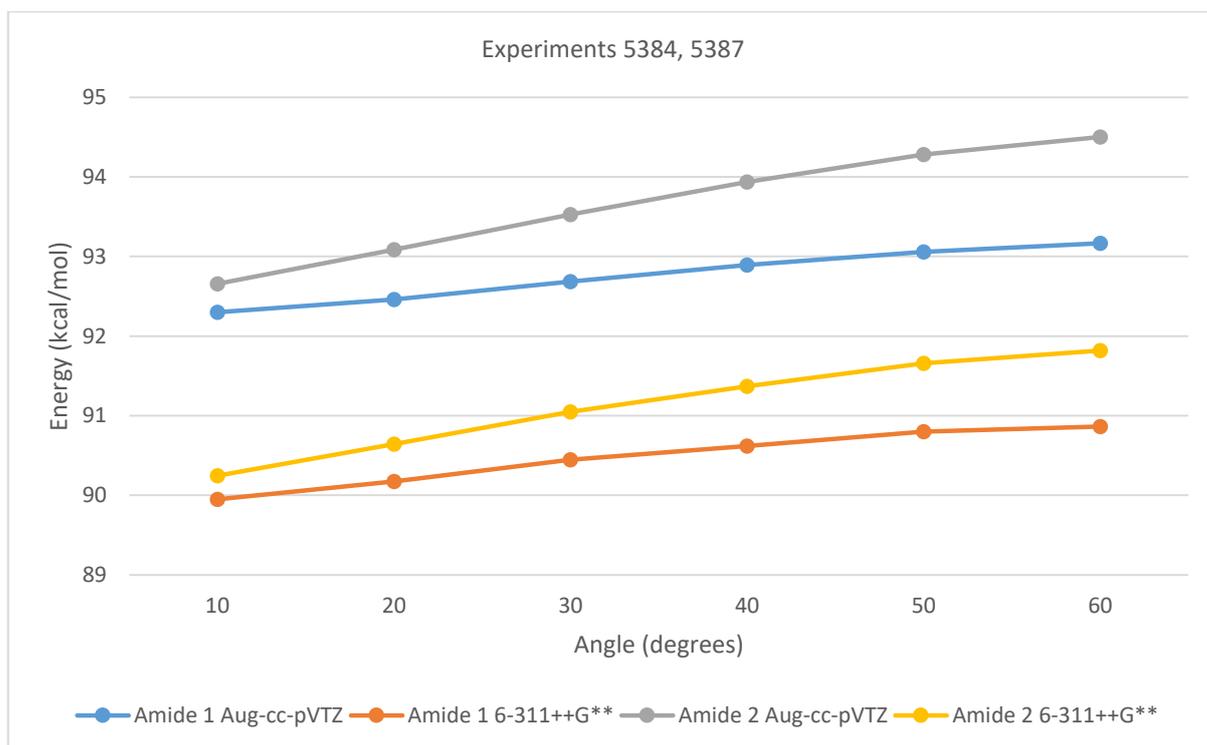

Figure 47. Amide Resonance Donor-Acceptor SOPT Minus Steric Exchange Energy for Hydrogen Bonded N-methylformamide 1 (C-O) and 2 (H-N) at Given C1-O1-N2 Angle and N1-C1-O1-N2 180 Degrees at LC-wPBE(w=0.4)/aug-cc-pVTZ

Table 6. Selected SOPT and Steric Exchange Energies for Single N-methylformamide

| Method | Basis | nlcopa | nlcop | olpcna | olpcn | olpchca | olpchc | olscos |
|---|---|---|---|---|---|---|---|---|
| B3LYP | Aug-cc-pVTZ | 67.81 | 16.78 | 25.91 | 13.22 | 22.7 | 13.56 | 42.86 |
| PBE1PBE | Aug-cc-pVTZ | 71.93 | 17.36 | 27.41 | 13.68 | 24.13 | 14.2 | 43.58 |
| CAM-B3LYP | Aug-cc-pVTZ | 87.58 | 18.35 | 30.3 | 14.2 | 27.15 | 14.64 | 42.99 |
| LC-wPBE | Aug-cc-pVTZ | 102.94 | 19.43 | 34.37 | 14.68 | 30.62 | 15.59 | 42.54 |
| B3LYP | 6-311++G** | 65.43 | 15.84 | 24.87 | 13.48 | 22.41 | 12.15 | 40.92 |
| PBE1PBE | 6-311++G** | 69.39 | 16.32 | 26.15 | 13.88 | 23.74 | 12.66 | 41.74 |
| CAM-B3LYP | 6-311++G** | 84.76 | 17.34 | 29.12 | 14.4 | 26.65 | 13.28 | 40.12 |
| LC-wPBE | 6-311++G** | 99.55 | 18.3 | 32.94 | 14.75 | 29.89 | 14.08 | 39.43 |

- nlcopa: N(lp)->C-O(p)* SOPT kcal/mol
- nlcop: N(lp)|C-O(p) Steric kcal/mol
- olpcna: O(lp-p)->C-N* SOPT kcal/mol
- olpcn: O(lp-p)->C-N Steric kcal/mol
- olpchca: O(lp-p)->C-HC* SOPT kcal/mol
- olpchc: O(lp-p)|C-HC Steric kcal/mol
- olscos: O(lp-s)|C-O(s) Steric kcal/mol

Table 7. Distribution of Hydrogen Bond C-O-N Angles Between Backbone Amides

| Series | 0 | 5 | 10 | 15 | 20 | 25 | 30 | 35 | 40 | 45 | 50 | 55 | 60 | 65 | 70 | 75 | 80 |
|---|---|---|---|---|---|---|---|---|---|---|---|---|---|---|---|---|---|
| banc-banc | 1196 | 3660 | 7555 | 11180 | 6400 | 1866 | 654 | 235 | 101 | 23 | 3 | | 2 | 1 | | | |
| banc-bac | 66 | 204 | 322 | 431 | 302 | 193 | 98 | 65 | 31 | 18 | 11 | 7 | 3 | 1 | 1 | | |
| bac-banc | 60 | 194 | 383 | 439 | 257 | 138 | 52 | 36 | 22 | 19 | 2 | 3 | 2 | 1 | 1 | | |
| bac-bac | 76 | 228 | 396 | 458 | 352 | 185 | 141 | 64 | 32 | 14 | 6 | 5 | 3 | 3 | 5 | | 1 |

- Banc: Backbone amide that has neither N-terminal nor C-terminal residue as random coil
- Bac: Backbone amide that has either N-terminal or C-terminal residue as random coil
- X-Y: Amide oxygen of X is hydrogen bonded to amide proton of Y

Hydrogen bond lengths are restricted to 2.2 angstroms. The hydrogen bonds reported are restricted N-C-O-N dihedral angles with absolute value less than or equal to 25 degrees or greater than or equal to 155 degrees and N-H-O angles less than or equal to 25 degrees. The numbers on columns of these tables refers to the start of 5 degree groupings of C-O-N angles. Small values intermittently exist for columns to the right of those shown.

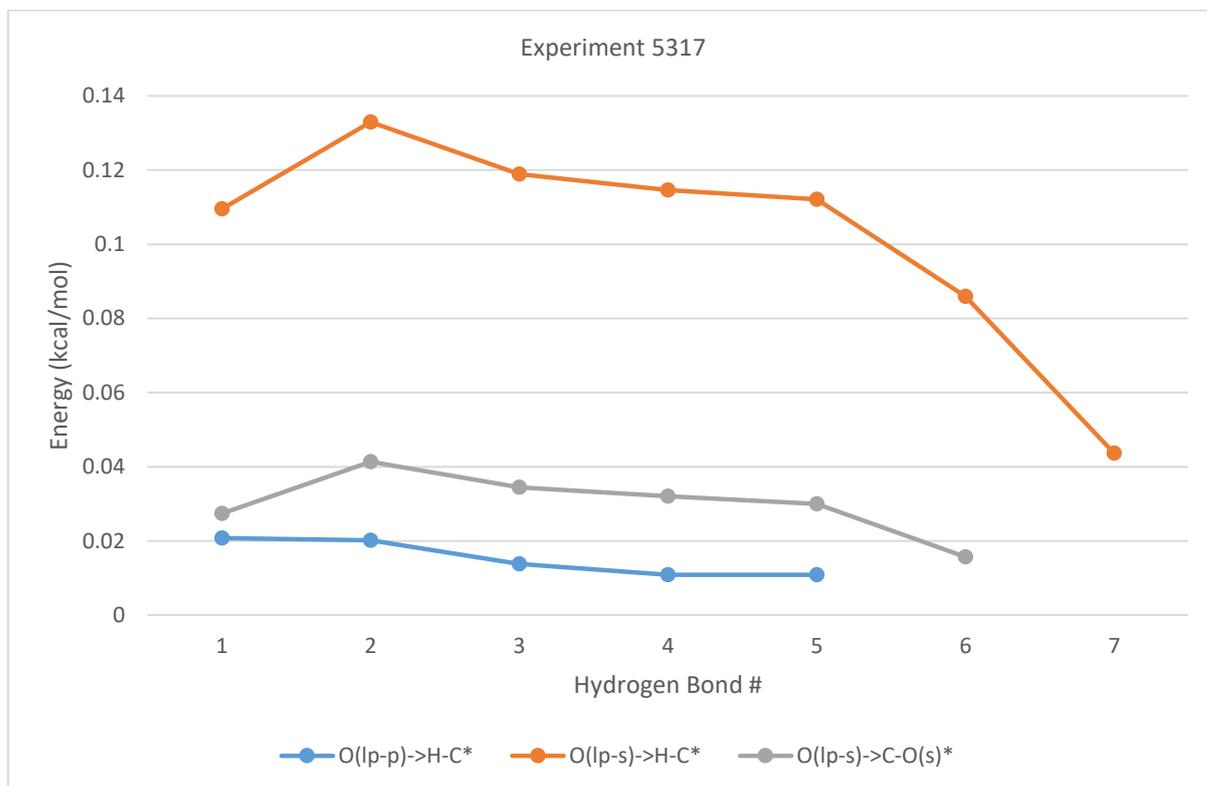

Figure 48. Donor-Acceptor SOPT Energy of Interactions Between Successive Units of Chain of 8 Formamides with Unconstrained Geometry Optimization at LC-wPBE(w=0.4)/6-311++G**

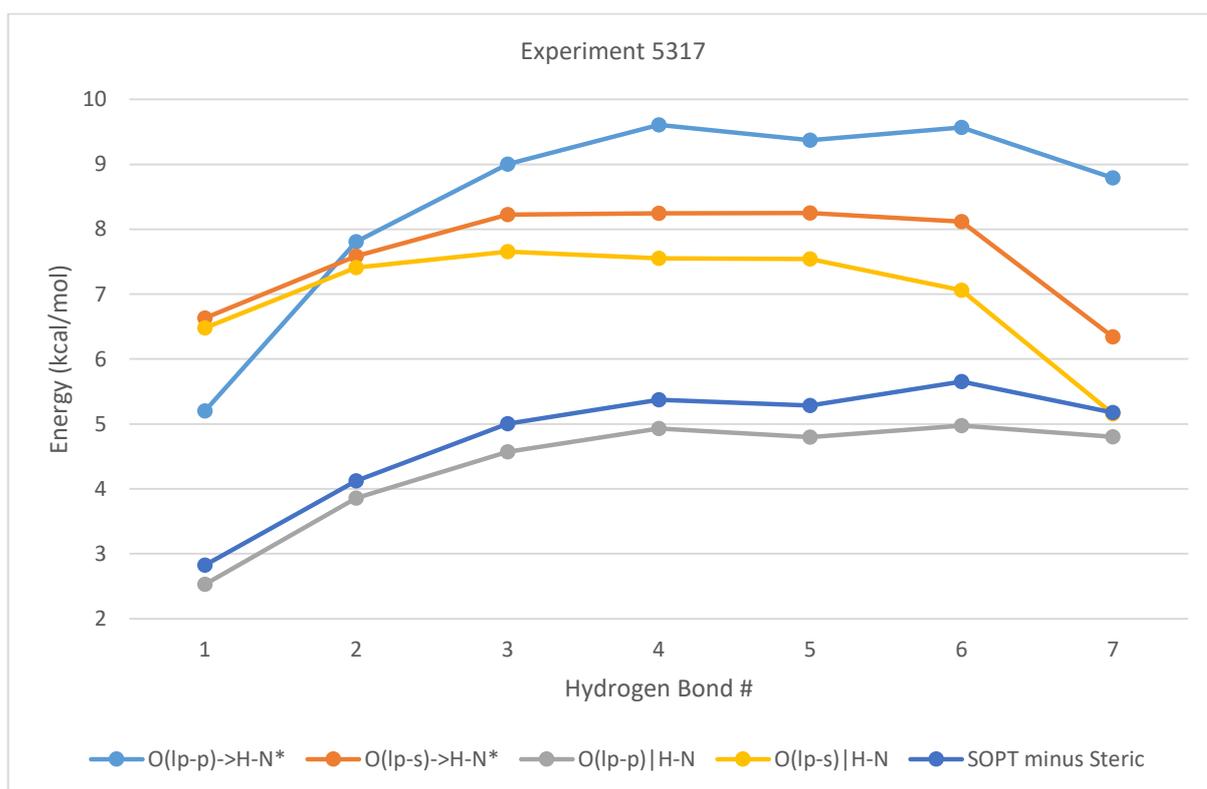

Figure 49. Hydrogen Bond Donor-Acceptor SOPT and Steric Exchange Energy in Chain of 8 Formamide Units with Unconstrained Geometry Optimization at LC-wPBE(w=0.4)/6-311++G**

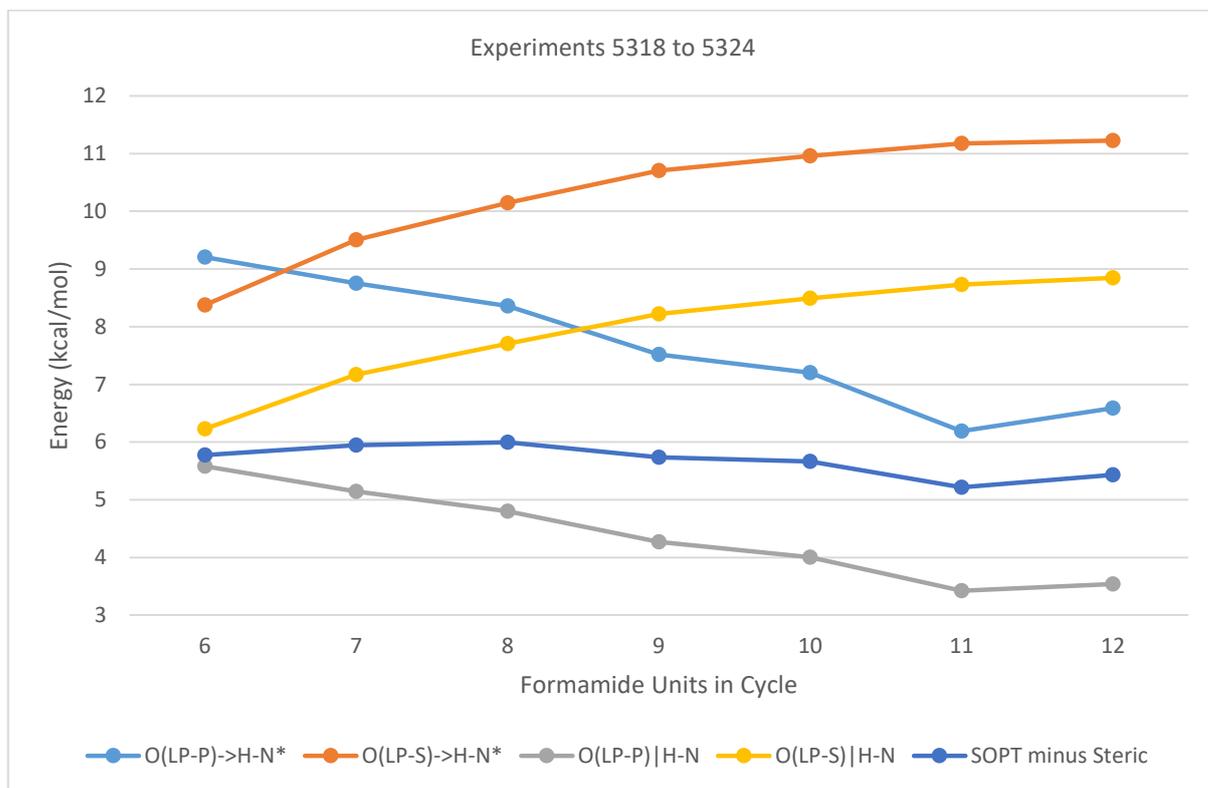

Figure 50. Donor-Acceptor SOPT and Steric Exchange Energy in Hydrogen Bonds of Planar Cycles of Formamide Optimized at LC-wPBE(w=0.4)/6-311++G**

## 10  Appendix 2. Polyvaline Parallel Beta Sheet

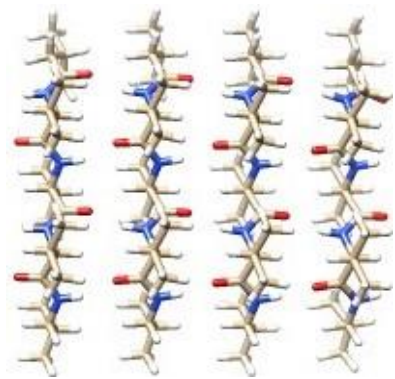

Figure 51. Polyvaline Parallel Beta Sheet. Experiment 997.

Table 8. N(lp)->C-O(p)* SOPT kcal/mol for Backbone Amides for Experiment 997

| ChainId | s1 | s2 | s3 | s4 |
|---|---|---|---|---|
| 1 | 47.84 | 48.63 | 49.76 | 75.07 |
| 2 | 113.87 | 117.74 | 94.94 | 113.08 |
| 3 | 62.81 | 100.44 | 105.27 | 106.41 |
| 4 | 112.34 | 118.61 | 110.07 | 105.92 |

Table 9. N(lp)->C-O(s)* SOPT kcal/mol for Backbone Amides for Experiment 997

| ChainId | s1 | s2 | s3 | s4 |
|---|---|---|---|---|
| 1 | 20.42 | 20.94 | 20.54 | 6.4 |
| 2 | 0 | 1.09 | 5.63 | 0.79 |
| 3 | 15.41 | 4.09 | 3.14 | 0.22 |
| 4 | 0.45 | 1.4 | 2.84 | 1.6 |

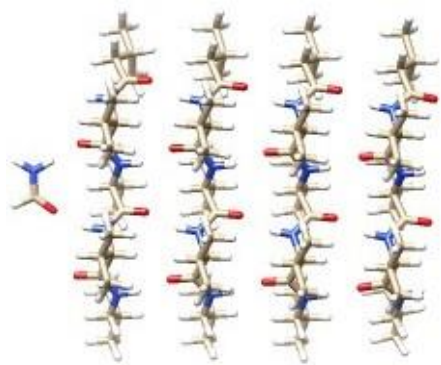

Figure 52. Polyvaline Parallel Beta Sheet Capped by One Formamide. Experiment 986.

Table 10. N(lp)->C-O(p)* SOPT kcal/mol for Backbone Amides for Experiment 986.

| ChainId | s1 | s2 | s3 | s4 |
|---|---|---|---|---|
| 1 | 49.11 | 46.55 | 51.84 | 73.82 |
| 2 | 116.82 | 111.79 | 108.28 | 101.46 |
| 3 | 101.43 | 104.65 | 100.28 | 104.43 |
| 4 | 109.7 | 117.59 | 110.91 | 111.87 |

Table 11. N(lp)->C-O(s)* SOPT kcal/mol for Backbone Amides for Experiment 986.

| ChainId | s1 | s2 | s3 | s4 |
|---|---|---|---|---|
| 1 | 19.35 | 22.37 | 19.21 | 6.79 |
| 2 | 2.23 | 2.3 | 2.58 | 2.28 |
| 3 | 5.2 | 3.45 | 4.1 | 0.82 |
| 4 | 0 | 1.28 | 2.68 | 0.7 |

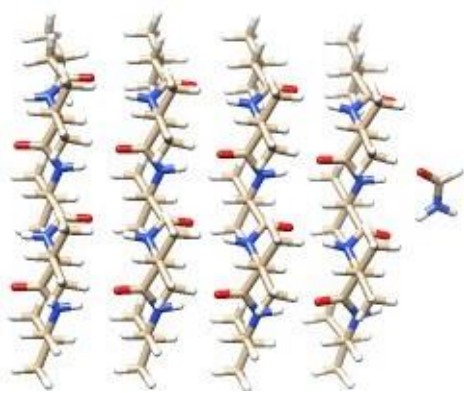

Figure 53. Polyvaline Parallel Beta Sheet Capped by One Formamide. Experiment 988.

Table 12. N(lp)->C-O(p)* SOPT kcal/mol for Backbone Amides for Experiment 988

| ChainId | s1 | s2 | s3 | s4 |
|---|---|---|---|---|
| 1 | 49.56 | 47.1 | 53.57 | 76.81 |
| 2 | 114.08 | 111.16 | 109.41 | 73.77 |
| 3 | 81.19 | 104.38 | 102.44 | 117.03 |
| 4 | 114.12 | 118.09 | 110.49 | 112.22 |

Table 13. N(lp)->C-O(s)* SOPT kcal/mol for Backbone Amides for Experiment 988

| ChainId | s1 | s2 | s3 | s4 |
|---|---|---|---|---|
| 1 | 19.55 | 21.89 | 17.87 | 6.13 |
| 2 | 0 | 2.25 | 2.35 | 11.21 |
| 3 | 8.75 | 3.23 | 3.55 | 0.5 |
| 4 | 0 | 1.39 | 2.74 | 0.56 |

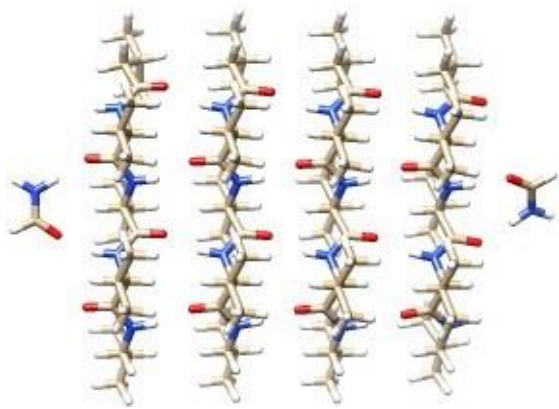

Figure 54. Polyvaline Parallel Beta Sheet with Two Capping Formamides. Experiment 998.

Table 14. N(lp)->C-O(p)* SOPT kcal/mol for Backbone Amides in Experiment 998.

| ChainId | s1 | s2 | s3 | s4 |
|---|---|---|---|---|
| 1 | 49.25 | 47.05 | 53.56 | 76.93 |
| 2 | 116.71 | 112.06 | 109.69 | 74.02 |
| 3 | 101.68 | 105.05 | 102.35 | 117.23 |
| 4 | 109.78 | 117.62 | 110.61 | 112.22 |

Table 15. N(lp)->C-O(s)* SOPT kcal/mol for Backbone Amides in Experiment 998

| ChainId | s1 | s2 | s3 | s4 |
|---|---|---|---|---|
| 1 | 19.23 | 21.9 | 17.93 | 6.14 |
| 2 | 2.26 | 2.26 | 2.32 | 11.14 |
| 3 | 5.14 | 3.33 | 3.65 | 0.5 |
| 4 | 0 | 1.25 | 2.65 | 0.55 |

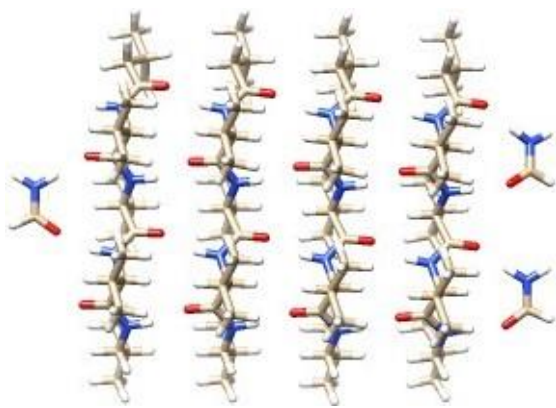

Figure 55. Polyvaline Parallel Beta Sheet with Three Capping Formamides. Experiment 991.

Table 16. N(lp)->C-O(p)* SOPT kcal/mol for Backbone Amides in Experiment 991.

| ChainId | s1 | s2 | s3 | s4 |
|---|---|---|---|---|
| 1 | 48.91 | 46.64 | 48.36 | 83.19 |
| 2 | 118.05 | 111.9 | 105.3 | 91.02 |
| 3 | 103.41 | 105.12 | 104.94 | 84.94 |
| 4 | 110.16 | 119.83 | 120.8 | 106.96 |

Table 17. N(lp)->C-O(s)* SOPT kcal/mol for Backbone Amides in Experiment 991.

| ChainId | s1 | s2 | s3 | s4 |
|---|---|---|---|---|
| 1 | 19.5 | 22.49 | 22.51 | 8.93 |
| 2 | 2.02 | 2.62 | 3.68 | 7.4 |
| 3 | 4.79 | 3.5 | 3.37 | 8.13 |
| 4 | 0 | 0.92 | 1.16 | 4.13 |

Table 18. Polyvaline Parallel Beta Sheet Totals

| exp | pact | sact | cts | ctp |
|---|---|---|---|---|
| 986 | 1520.52 | 95.33 | 105.38 | 23.21 |
| 998 | 1515.8 | 100.25 | 106.91 | 23.32 |
| 991 | 1509.52 | 115.16 | 108.22 | 25.04 |
| 988 | 1495.4 | 101.97 | 106.87 | 23.16 |
| 997 | 1482.82 | 104.97 | 105.69 | 23.41 |

- exp: experiment number
- pact: total primary backbone amide charge transfer N(lp)->C-O(pi)* kcal/mol
- sact: total secondary backbone amide charge transfer N(lp)->C-O(sigma)* kcal/mol
- cts: total intra beta sheet inter amide HB charge transfer O(lp-s)->H-N* kcal/mol
- ctp: total intra beta sheet inter amide HB charge transfer O(lp-p)->H-N* kcal/mol

## 11 Appendix 3. Polyvaline Antiparallel Beta Sheet

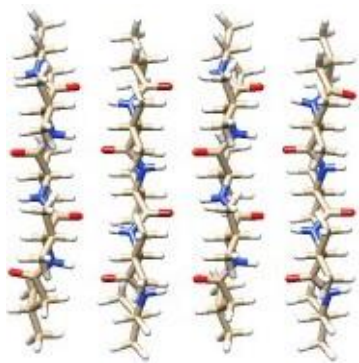

Figure 56. Polyvaline Antiparallel Beta Sheet. Experiment 995.

Table 19. N(lp)->C-O(p)* SOPT kcal/mol for Experiment 995.

| ChainId | s1 | s2 | s3 | s4 |
|---|---|---|---|---|
| 1 | 112.59 | 57.1 | 123.58 | 78.94 |
| 2 | 109.31 | 125.38 | 123.8 | 95.55 |
| 3 | 92.98 | 123.15 | 123.07 | 110.03 |
| 4 | 80.79 | 123.08 | 59.46 | 113.03 |

Table 20. N(lp)->C-O(s)* SOPT kcal/mol for Experiment 995.

| ChainId | s1 | s2 | s3 | s4 |
|---|---|---|---|---|
| 1 | 0 | 17.84 | 0.13 | 5.31 |
| 2 | 0.5 | 0.51 | 0.37 | 4.03 |
| 3 | 4.67 | 0.62 | 0.96 | 0.39 |
| 4 | 4.82 | 0.18 | 16.62 | 0 |

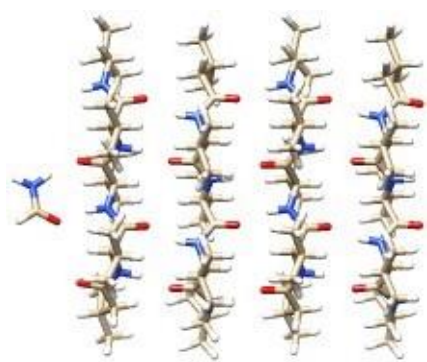

Figure 57. Polyvaline Antiparallel Beta Sheet with One Capping Formamide. Experiment 989.

Table 21. N(lp)->C-O(p)* SOPT kcal/mol for Experiment 989.

| ChainId | s1 | s2 | s3 | s4 |
|---|---|---|---|---|
| 1 | 112.23 | 60.87 | 123.99 | 82.14 |
| 2 | 120.32 | 128.32 | 123.54 | 108.38 |
| 3 | 97.82 | 124.81 | 125.76 | 110.8 |
| 4 | 84.52 | 121.66 | 65.35 | 113.24 |

Table 22. N(lp)->C-O(s)* SOPT kcal/mol for Experiment 989

| ChainId | s1 | s2 | s3 | s4 |
|---|---|---|---|---|
| 1 | 0 | 16.05 | 0 | 4.54 |
| 2 | 0.2 | 0 | 0.13 | 1.15 |
| 3 | 4.4 | 0.43 | 0.47 | 0.19 |
| 4 | 4.05 | 0.2 | 13.83 | 0 |

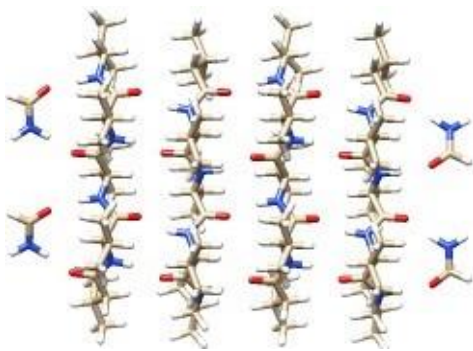

Figure 58. Polyvaline Antiparallel Beta Sheet with Four Capping Formamides. Experiment 990.

Table 23. N(lp)->C-O(p)* SOPT kcal/mol for Backbone Amides for Experiment 990.

| ChainId | s1 | s2 | s3 | s4 |
|---|---|---|---|---|
| 1 | 123.15 | 63.09 | 125.92 | 69.07 |
| 2 | 98.09 | 125.71 | 123.39 | 109.07 |
| 3 | 113.94 | 122.95 | 122.72 | 106.29 |
| 4 | 66.31 | 125.16 | 67.18 | 122.85 |

Table 24. N(lp)->C-O(s)* SOPT kcal/mol for Backbone Amides for Experiment 990.

| ChainId | s1 | s2 | s3 | s4 |
|---|---|---|---|---|
| 1 | 0.46 | 15.56 | 0 | 12.69 |
| 2 | 4.8 | 0.26 | 0.2 | 2.62 |
| 3 | 1.77 | 0.46 | 0.87 | 3.14 |
| 4 | 13.67 | 0 | 13.74 | 0.49 |

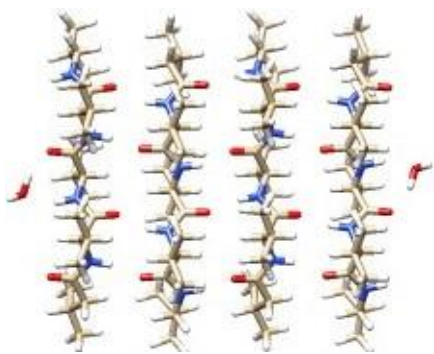

Figure 59. Polyvaline Antiparallel Beta Sheet with Two Capping Waters. Experiment 994.

Table 25. N(lp)->C-O(p)* SOPT kcal/mol for Backbone Amides for Experiment 994.

| ChainId | s1 | s2 | s3 | s4 |
|---|---|---|---|---|
| 1 | 110.03 | 64.83 | 120.29 | 85.32 |
| 2 | 105.91 | 120.88 | 114.94 | 87.21 |
| 3 | 81.91 | 114.63 | 114.03 | 107.39 |
| 4 | 85.25 | 120.18 | 72.08 | 111.67 |

Table 26. N(lp)->C-O(s)* SOPT kcal/mol for Backbone Amides for Experiment 994.

| ChainId | s1 | s2 | s3 | s4 |
|---|---|---|---|---|
| 1 | 0.41 | 14.2 | 0.53 | 3.42 |
| 2 | 3.26 | 1.61 | 2.43 | 7.89 |
| 3 | 9.33 | 2.56 | 2.91 | 3 |
| 4 | 3.47 | 0.59 | 11.41 | 0.17 |

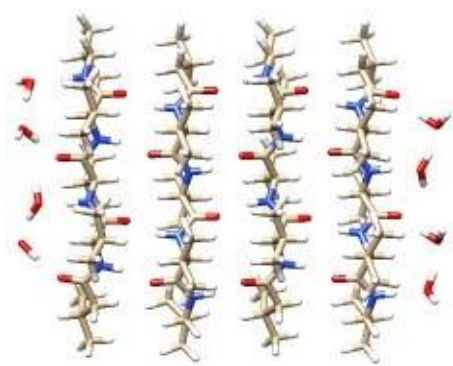

Figure 60. Polyvaline Antiparallel Beta Sheet with Eight Capping Waters. Experiment 996.

Table 27. N(lp)->C-O(p)* SOPT kcal/mol for Backbone Amides for Experiment 996.

| ChainId | s1 | s2 | s3 | s4 |
|---|---|---|---|---|
| 1 | 125.37 | 66.77 | 124.07 | 65.25 |
| 2 | 91.69 | 119.48 | 123.81 | 103.46 |
| 3 | 103.8 | 123.92 | 118.36 | 107.95 |
| 4 | 72.62 | 123.9 | 71.69 | 113.72 |

Table 28. N(lp)->C-O(s)* SOPT kcal/mol for Backbone Amides for Experiment 996.

| ChainId | s1 | s2 | s3 | s4 |
|---|---|---|---|---|
| 1 | 0.37 | 14.03 | 0.32 | 14.28 |
| 2 | 9.63 | 1.62 | 0 | 4.46 |
| 3 | 3.94 | 0 | 1.83 | 5.04 |
| 4 | 11.34 | 0.38 | 12.03 | 2.46 |

Table 29. Polyvaline Antiparallel Beta Sheet Totals

| exp | pact | sact | cts | ctp |
|---|---|---|---|---|
| 989 | 1703.74 | 45.63 | 100.2 | 46.71 |
| 990 | 1684.9 | 70.74 | 106.1 | 49.95 |
| 996 | 1655.88 | 81.74 | 105.5 | 49.62 |
| 995 | 1651.84 | 56.95 | 98.77 | 47.32 |
| 994 | 1616.54 | 67.21 | 104.3 | 44.67 |

- exp: experiment number
- pact: total primary backbone amide charge transfer N(lp)->C-O(pi)* kcal/mol
- sact: total secondary backbone amide charge transfer N(lp)->C-O(sigma)* kcal/mol
- cts: total intra beta sheet inter amide HB charge transfer O(lp-s)->H-N* kcal/mol
- ctp: total intra beta sheet inter amide HB charge transfer O(lp-p)->H-N* kcal/mol